\documentclass[
aps,%
11pt,%
final,%
notitlepage,%
showkeys,%
oneside,%
onecolumn,%
nobibnotes,%
nofootinbib,%
superscriptaddress,%
noshowpacs,%
centertags]{revtex4-2}

\input maik.sty 
\def\refitem#1{\relax}

\newcommand{\micron}{~$\mu$m}
\newcommand{\arcsec}{^{\prime\prime}}
\newcommand{\arcmin}{^{\prime}}
\newcommand{\pcm}{~cm$^{-2}$}	
\newcommand{\pcmm}{~cm$^{-3}$}	
\newcommand{\kms}{~km\,s$^{-1}$}
\newcommand{\msun}{~$M_\odot$}
\newcommand{\Hii}{H~{\sc ii}}

\begin{document}

\title{STUDY OF THE PHYSICAL AND CHEMICAL PROPERTIES OF DENSE CLUMPS AT DIFFERENT EVOLUTIONARY STAGES IN SEVERAL REGIONS OF MASSIVE STAR AND STELLAR CLUSTER FORMATION}

\author{\firstname{A.~G.}~\surname{Pazukhin}}
\email{a.pazuhin@ipfran.ru}
\author{\firstname{I.~I.}~\surname{Zinchenko}}
\email{zin@ipfran.ru}
\affiliation{Federal Research Center A.V. Gaponov-Grekhov Institute of Applied Physics of the Russian Academy of Sciences, Nizhny Novgorod 603950, Russia}
\affiliation{National Research Lobachevsky State University of Nizhny Novgorod, Nizhny Novgorod 603950, Russia}

\author{\firstname{E.~A.}~\surname{Trofimova}}
\email{tea@ipfran.ru}
\affiliation{Federal Research Center A.V. Gaponov-Grekhov Institute of Applied Physics of the Russian Academy of Sciences, Nizhny Novgorod 603950, Russia}

\begin{abstract}
 
Massive stars play an important role in the Universe. Unlike low-mass stars, the formation of these objects located at great distances is still unclear. It is expected to be governed by some combination of self-gravity, turbulence, and magnetic fields.
In this work, we aim to study the chemical and physical conditions of dense clumps at different evolutionary stages.
We performed observations towards 5 regions of massive star and stellar cluster formation (L1287, S187, S231, DR~21(OH), NGC~7538) with the IRAM-30m telescope. We covered the 2 and 3--4~mm wavelength bands and analysed the lines of HCN, HNC, HCO$^+$, HC$_3$N, HNCO, OCS, CS, SiO, SO$_2$, and SO.  Using astrodendro algorithm on the 850\micron\ dust emission data from the SCUBA Legacy catalogue, we determined the masses, H$_2$ column densities, and sizes of the clumps. Furthermore, the kinetic temperatures, molecular abundances, and dynamical state were obtained. The Red Midcourse Space Experiment Source survey (RMS) was used to determine the clump types.
A total of 20 clumps were identified. Three clumps were found to be associated with the \Hii\ regions, 10 with young stellar objects (YSOs), and 7 with submillimetre emission.
The clumps have typical sizes of about 0.2~pc and masses ranging from 1 to $10^{2}$\msun, kinetic temperatures ranging from 20 to 40~K and line widths of $\rm H^{13}CO^{+} (1-0)$ approximately 2~$\rm km\,s^{-1}$. We found no significant correlation in the line width--size and the line width--mass relationships. However, a strong correlation is observed in mass--size relationships. The virial analysis indicated that three clumps are gravitationally bound. Furthermore, we suggested that magnetic fields of about 1~mG provide additional support for clump stability. The molecular abundances relative to H$_2$ are approximately $10^{-10}-10^{-8}$.

\end{abstract}

\keywords{star formation, interstellar medium, molecular clouds, interstellar molecules, astrochemistry}

\maketitle

\section{Introduction}

In spite of their short lifetime, high-mass stars (also OB stars, $L>10^3\, L_\odot$, $M>8$\msun) play an important role in the Universe. Their formation and evolution are still poorly understood (e.g.,~\cite{Motte18}). According to the review in~\cite{Zinnecker07}, there are several evolutionary stages. Objects associated with the first phase of high-mass star formation are called IR dark clouds (IRDCs), having dense and cold gas likely represent the initial conditions of high-mass star formation. Hot molecular cores (HMCs) have large masses of warm and dense gas, large abundances of complex organic and maser emission. Finally, ionising radiation from the embedded stellar population disrupts the parent molecular cloud, and an ultra-compact \Hii\ region (UCH{\sc ii}) is formed.

In the pioneering work \cite{Larson81}, about 50 giant molecular clouds (GMCs) were observed to investigate the effect of turbulence on gravitational collapse. The obtained fundamental relations between the line width, cloud mass, density, and size are the subject of active research and discussion (e.g.,~\cite{Solomon87,Fuller92,Caselli95,Zinchenko2000,Heyer04,Traficante18}).
Furthermore, A significant issue is the interaction between the magnetic field and the self-gravity of the cloud (e.g.,~\cite{Bertoldi92,Kauffmann13,Pillai15}).

The chemical evolution has been discussed to understand the conditions in massive star-forming regions. A number of recent works have identified variations in molecular abundances with evolutionary stages, from IRDCs to \Hii\ regions (e.g.,~\cite{Vasyunina11,Sanhueza12,Gerner14,Rathborne16,Urquhart19}). 

{In this work, we analyse observational data obtained in 2019 with the IRAM-30m telescope. In the previous works based on the same data set we investigate the intensity ratio of HCN and HNC as a temperature indicator~\cite{Pazukhin22}, and analyse the variation of deuterium fractionation in HCO$^{+}$, HCN, HNC, N$_2$H$^+$, and NH$_3$ molecules depending on temperature and density~\cite{Pazukhin23}. Here our goal is an investigation of the physical and chemical properties of dense clumps at different evolutionary stages. The clumps are identified by dust emission at 850\micron\ from the SCUBA catalogue~\cite{scuba}. Their type is determined using the Red MSX Source (RMS) catalogue~\cite{Lumsden13}. 
 We investigate well-known regions of massive star and stellar cluster formation (L1287, S187, S231, DR~21(OH), NGC~7538), but this analysis reveals some new details about their structure and properties. We note that water and methanol masers, and a relatively powerful IR source are observed in L1287. However, no massive protostar has been detected.  Evidences suggest the presence of a massive clump and the formation of a star cluster (see, e.g., \cite{Sepulveda20}).}

The paper is organised as follows. Sections \ref{obs} and \ref{datared} describe the observations and data reduction. An analysis of molecular line spectra and clumps is presented in Sect. \ref{results}. We discuss the Larson relationships and abundance estimates in Sect. \ref{diss}. A summary of the main conclusions is presented in Sect. \ref{sum}.

\section{Observational data}\label{obs}

In September 2019, with the 30-m radio telescope of the Institut de Radioastronomie Millim{\'e}trique (IRAM), we observed five massive star forming regions at wavelengths of 2 and 3--4~mm (in the framework of the project 041-19). The list of sources is given in Table~\ref{tab:source}. Table~\ref{tab:lines} contains the list of the observed molecular lines with some spectroscopic parameters. Transition frequencies and upper level energies are taken from The Cologne Database for Molecular Spectroscopy (CDMS)\footnote{\url{http://cdms.de}} \cite{cdms}.

The full beam width at half maximum at the observed frequencies ranged from $38\arcsec$ to $18\arcsec$.  Antenna temperatures $T^*_{\rm A}$ were converted to the main beam brightness temperature $T_{\rm mb}$, using the main beam efficiency $B_{\rm eff}$, which was determined by the Ruze's formula in accordance with the IRAM recommendations\footnote{\url{https://publicwiki.iram.es/Iram30mEfficiencies}} and ranged from 0.73 to 0.83. The minimum system noise temperatures were approximately 100~K in the 3~mm range and approximately 200~K in the 2~mm range.

Observations were carried out in the On-The-Fly (OTF) mode over a mapping area of a $200\arcsec\times200\arcsec$ in the total power mode. The reference position was chosen with a shift of $10\arcmin$ in right ascension. In some extended sources, that is DR~21(OH) and NGC~7538, two partially overlapping areas were observed. The pointing accuracy was checked periodically by observations of nearby continuum sources.

\section{Data reduction}\label{datared}

The GILDAS/CLASS software\footnote{\url{http://www.iram.fr/IRAMFR/GILDAS}} was used for the data reduction. The spectra were fitted with Gaussian profiles using the \texttt{LMFIT} package \cite{lmfit}.  In the analysis, integrated intensity was obtained from the Gaussian profile area with their errors as the fitting errors.  For the spectra with hyperfine structure we assume that the widths of all components are equal, and the spacings between them are known. We note that two velocity components at about $-4$ and 0\kms\ are observed in the source DR~21(OH) (see details in~\cite{Schneider}). In the reduction, the components were separated, and only the $-4$\kms\ component has been used for the analysis, since it is stronger and is detected throughout the source.

\section{Results}\label{results}

\subsection{Clumps identification from dust emission}

To extract the clumps, we use Python Astrodendro\footnote{\url{http://www.dendrograms.org}}~\cite{Rosolowsky08}. A dendrogram is employed to represent a hierarchical data structure. A dendrogram consists of two types of structures: branches, which are structures which split into multiple sub-structures, and leaves, which are structures that have no sub-structure and represent the final structures. In astrodendro methods, three key parameters must be set: \texttt{min\_value} (initial value), \texttt{min\_delta} (minimum height of the leaf to consider it independent) and \texttt{min\_npix} (minimum number of pixels of the leaf to consider it independent). We set min\_value = 4$\sigma$, where $\sigma$ is the mean error in $\rm Jy\,beam^{-1}$ for dust emission data, min\_delta = $\sigma$, and min\_npix = FWHM (22.9$\arcsec\approx4$~pix for the SCUBA Legacy catalogue at 850\micron). We define the leaves on the dendrogram as clumps. 
{To determine the clump flux, we sum the pixel brightness within the clump. If leaves are located on a branch of the dendrogram, we subtract the brightness values of that branch (e.g.,~\cite{Ragan13}).}

Figure~\ref{fig:clump} shows maps of 850\micron\ dust emission from the SCUBA Legacy catalogue. The clumps obtained using the astrodenro algorithm are also shown. A total of 20 clumps were identified across the five sources.
 The clump classification was based on the results of the Red MSX Source (RMS) survey~\cite{Lumsden13}. The RMS survey is a mid-infrared selection of young massive protostars and \Hii\ regions, identified by the MSX. Three clumps were found to be associated with the \Hii\ regions, 10 with YSOs. Seven clumps are visible at submillimetre emission but are dark in the MIR. We designate these clumps as "submm". Furthermore, the maser database\footnote{\url{https://maserdb.net}} \cite{Ladeyschikov19}  indicates the existence of water, hydroxyl, and methanol masers in clumps associated with YSO and \Hii\ regions. The list of identified clumps is summarised in Table~\ref{tab:clumps}. Three-colour WISE images of sources are presented in Appendix (see Fig.~\ref{fig:wise}).

\subsection{Moment maps and spectra}

Figures~\ref{fig:spectra} and~\ref{fig:add_spectra} include the average spectra for the clumps. As mentioned above, Table~\ref{tab:lines} contains the list of the observed molecular lines. Most of these molecules (e.g., HCN, HNC, HCO$^+$) are known to be good tracers of dense gas (e.g.,~\cite{Vasyunina11}). Furthermore, HNCO, SO, and SiO are effective tracers of shocked gas. The HCO$^+$ and SiO are also good tracers of outflows. 
For all clumps, the HNC, HCO$^+$, and HCN lines show strong emission and clear detection, but they mainly exhibit a red-shifted self-absorption. For YSOs and \Hii\ regions, the existence of wings is found in the HCO$^+$ and SiO spectra. OCS shows rather weak emission and is detected in less than half of the clumps. 
The detection rates of molecular lines in clumps with YSO and \Hii\ regions are higher than that in submm.
The average line widths are about 3\kms, with a minimum for C$^{34}$S (2.4\kms) and a maximum for SiO (5.5\kms). In general, line widths increase with evolutionary type.

The optically thin dense gas tracer, H$^{13}$CO$^{+}$(1--0), is employed to analyse the kinematics and dynamics of the gas. Figures~\ref{fig:moment} and~\ref{fig:add_moment} illustrate moment maps obtained by fitting a Gaussian profile. The obtained clumps are relatively warm (>20~K), so the depletion effect is not significant for H$^{13}$CO$^{+}$. We found a correlation between this tracer and the dust emission. 

\subsection{Clump masses from dust emission}
To estimate the clump masses, the following equation was used \cite{Kauffmann08}:
\begin{equation}
  \begin{array}{l}
    M = 
    \displaystyle 0.12 \, M_{\odot}
    \left( {\rm e}^{14.39 (\lambda / {\rm mm})^{-1}
        (T / {\rm K})^{-1}} - 1 \right) 
     \left( \frac{\kappa_{\nu}}{\rm cm^2\,g^{-1}} \right)^{-1}\\
    \quad \displaystyle
    \times \left( \frac{F_{\nu}}{\rm Jy} \right)
    \left( \frac{d}{\rm 100 ~pc} \right)^2
    \left( \frac{\lambda}{\rm mm} \right)^{3}
  \end{array}
\end{equation}
where the gas-to-dust ratio is assumed to be 100,
$\lambda$~is the wavelength,
$F_{\rm\nu}$~is the integrated flux, 
$T_{\rm dust}$~is the dust temperature,
$d$~is the distance to the object.
Dust opacity $\kappa_{\rm \nu}=1.82$~cm$^2$g$^{-1}$ at 850\micron~\cite{Ossenkopf94}.
The dust temperature $T_{\rm dust}$ was assumed to be 20~K.  
If leaves are located on a branch of the dendrogram, the clump flux $F_{\rm\nu}$ is determined after subtracting the brightness value of the parent branch $F_{\rm bg}$ (see Table~\ref{tab:clumps}).
Thus, the H$_2$ volume density is derived as follows:
\begin{equation}
    n({\rm H_2}) = \frac{M}{\mu_{\rm H_{2}} m_{\rm  H} (4/3\pi R_{\rm eff}^3)},
\end{equation}
where $m_{\rm H}$~is the hydrogen mass, the mean molecular weight $\mu_{\rm H_2}=2.8$~\cite{Kauffmann08}, $R_{\rm eff}$ is the effective radius determined from the clump area $\sqrt{A/\pi}$. 
 
\subsection{Kinetic temperature}

The HCN-to-HNC ratio is strongly dependent on the kinetic temperature \cite{Hirota}. In~\cite{Hacar20} it is proposed to utilise the $J=1-0$ HCN and HNC intensity ratio as a temperature indicator. We adopted the kinetic temperature maps derived from the integrated intensity ratios of the $J=1-0$ HCN and HNC and their $^{13}$C isotopologues from \cite{Pazukhin23}. The kinetic temperatures range from 20~K (S187) to 40~K (NGC~7538).

\subsection{Virial parameter}
The virial parameter, which compares the virial mass to the gas mass, provides one method to investigate the cloud stability. Low values are observed in high-mass star-forming regions. This may indicate a rapid collapse compared to low-mass star formation \cite{Kauffmann13}.  The virial parameter is defined as follows \cite{Bertoldi92}:
\begin{equation}
    \alpha_{\rm vir}=\frac{5\sigma_{\rm tot}^2R_{\rm eff}}{GM},
    \label{eq:alpha}
\end{equation}
where $G$~is the gravitational constant. For $\alpha_{\rm vir}<2,$ the self-gravity is a significant factor, assuming a weak impact of other forces. The dynamic state of the cloud is affected by the magnetic field and the external pressure, although this is difficult to observe.  Thus, the virial parameter is one of the simple approximations to determine the cloud state.  

The virial parameter includes the total velocity dispersion ($\sigma_{\rm tot}$), which can be estimated from observed line width ($\sigma_{\rm {\rm obs}}=\Delta \upsilon/\sqrt{8\, {\rm ln}\, 2}$) considering the thermal and non-thermal component:
\begin{equation}
    \sigma_{\rm tot}^2=\sigma_{\rm {\rm nt}}^2+c_{\rm s}^2=\sigma_{\rm obs}^2-\frac{k_{\rm B}T_{\rm kin}}{\mu_{\rm mol} m_{\rm p}}+{\frac{k_{\rm B}T_{\rm kin}}{\mu_{\rm p} m_{\rm p}}},
    \label{eq:sigmatot_definition}
\end{equation}
where the mean molecular weight per free particle $\mu_{\rm p} = 2.37$, $\mu_i$~is the molecular weight for the observed species, $m_{\rm p}$~is the proton mass, $k_{\rm B}$~is the Boltzmann constant and $T_{\rm kin}$ is the kinetic temperature. In addition, the Mach number, $\mathcal{M}=\sigma_{\rm nt}/c_{\rm s}$, was obtained as an indicator of turbulence in the cloud.

\subsection{Column density of molecules}

The column density for the optically thin, Rayleigh-Jeans and negligible background approximations is calculated as follows~\cite{Mangum15}:

\begin{equation} 
N_{\rm tot} = \frac{3 k_{\rm B}}{8 \pi^3 \nu S \mu^2} \frac{Q_{\rm rot}}{g_{\rm u}} \exp\left(\frac{E_{\rm u}}{k T_{\rm ex}}\right) \int T_{\rm R} d\upsilon,
\end{equation} 
where $\int T_{\rm R}d\upsilon$ is the integrated intensity of the line in K\kms, $E_{\rm u}$ is the energy of the upper level, $S$ is the strength of the line, $\mu$ is the dipole moment, $\nu$ is the transition frequency, $g_u$ is the statistical weight of the upper level, $Q_{\rm rot}$ is the partition function, the excitation temperature $T_{\rm ex}$ was assumed to be equal to the kinetic $T_{\rm kin}$, and the brightness temperature $T_{\rm R}$ equals to the main beam temperature $T_{\rm mb}$. To calculate column density of molecules, we use Python Pyspeckit\footnote{\url{https://pyspeckit.readthedocs.io}}~\cite{Ginsburg22}.
For the optically thick lines of HCN, HNC, and HCO$^+$, their optically thin isotopologues containing $^{13}$C were used. C$^{34}$S was employed to derive the column density of CS. To do this, the column density of optically thin isotopologues was multiplied by a factor derived from the isotope ratio \cite{Yan23}: 
\begin{eqnarray}
    \frac{\rm ^{12}C}{\rm  ^{13}C}= 4.77 \times R_{\rm GC} + 20.76,\\
    \frac{\rm  ^{32}S}{\rm  ^{34}S} = 0.75 \times R_{\rm GC} + 15.52,
\end{eqnarray}
where R$_{\rm GC}$ is the galactocentric distance in kpc.

\section{Discussion}\label{diss}

\subsection{Larson relations}
In~\cite{Larson81} the relations between linewidth, cloud mass, volume density, and cloud size were derived. The Larson's laws are as follows: (1) the velocity dispersion as a function of the cloud size, $\sigma \propto R^{0.38}$, (2)  the velocity dispersion as a function of the cloud mass, $\sigma \propto M^{0.2}$, and (3) the mean density as a function of the cloud size, $n \propto R^{-1.1}$. Finally, the mass-size relationship can be defined from (1) and (2), $M \propto R^{1.9}$.

 The list of clump parameters is summarised in Tables~\ref{tab:clumps} and~\ref{tab:params}. In Fig.~\ref{fig:size}, we show the relationship of $\sigma_{\rm obs}-R_{\rm eff}$, $n({\rm H_2})-R_{\rm eff}$ and $M-R_{\rm eff}$. Figure~\ref{fig:mass} presents the $\sigma_{\rm obs}-M$ and  $\alpha_{\rm vir}-M$ relations.
The results indicate a weak correlation between the $\sigma_{\rm obs}$ and $R_{\rm eff}$, as evidenced by the Spearman's rank correlation coefficient\footnote{\texttt{scipy.stats.spearmanr}} $r_{\rm s}=0.16$ (p-value = 0.510). 
The correlation between $\sigma_{\rm obs}$ and $M$ is also weak ($r_{\rm s}=0.4$, p-value = 0.081). 
We found no significant correlation between  $n(\rm H_2)$ and $L$ ($r_{\rm s}=-0.2$, p-value = 0.394).
The mass and size present a strong correlation ($r_{\rm s}=0.9$, p-value = $6.2\times10^{-5}$), with a power-law index of $3.3\pm0.36$.

The Larson relations are extensively studied in various samples of star-forming regions. We note that these relations may vary depending on the observation resolution, the object type, and the methods used to determine the mass and size, as well as the line widths of the molecular tracers. 

 In~\cite{Larson81} the velocity dispersion was shown as a function of the cloud size for GMCs, $\sigma \propto R^{0.38}$. Later, these results were modified to $\sigma\propto R^{0.5}$ \cite{Heyer04}, which is comparable to the results in~\cite{Solomon87}. In the paper~\cite{Fuller92} based on CS, C$^{18}$O, and NH$_{3}$ a significant correlation between line width and radius, $\Delta \upsilon \propto R^{0.5}$, was found. In addition, in~\cite{Caselli95} a trend $\Delta \upsilon_{NT} \propto R^{0.21}$ was obtained for massive dense cores in Orion, which contrasts with their findings for low-mass cores where the slope is 0.53. These results are consistent with the slopes found for the massive dense cores observed in CS and C$^{34}$S~\cite{Zinchenko2000}, and in $\rm N_{2}H^{+}$~\cite{Pirogov03}. In~\cite{Traficante18} a sample of 213 massive clumps was studied, and a weak correlation between velocity dispersion and radius, with a slope of $0.1$ was found. 

Figure~\ref{fig:size}c demonstrates mass--size relationships and magnetic fields. In~\cite{Crutcher12} an upper limit for the magnetic field strength was found, which increases with increasing density as $B =B_{0}\, (n/10^4\,{\rm cm^{-3}})^{0.65}$ (the red dashed lines in Fig.~\ref{fig:size}c). For $B_{0}=150\, \mu$G and a mean density estimate of approximately 10$^5$\pcmm, the Crutcher's relation suggests $B\sim 1$~mG. The clumps with masses over 100\msun are situated within a magnetic field region ($B>0\, \mu$G), and exhibit a low virial parameter ($\alpha_{\rm vir}<2$; Fig.~\ref{fig:mass}b). The density and mass of these clouds are such that the thermal pressure and random gas motions are insufficient to provide significant support against self-gravity. This is evidenced by a low virial parameter, which suggests that magnetic fields provide additional support. We note that the low virial parameter may be biased by calculation methods, leading to uncertainties of a factor of two or more (see, e.g., \cite{Kauffmann13,Singh21}).

The mass--radius relation can also be a useful tool to investigate clumps that are likely to form high-mass stars, following the empirical threshold ($M\geq870\, R^{1.33}$)~\cite{Kauffmann10}.
Clumps "2"\ and "3"\ in NGC~7538 and "2"\ in DR~21(OH) are located above this threshold (Fig.~\ref{fig:size}c).

\subsection{Molecular abundances relative to H$_2$}

We have obtained the abundances of HCN, HNC, HCO$^+$, HC$_3$N, HNCO, OCS, CS, SiO, SO$_2$ and SO molecules. In cases where multiple rotational transitions of a molecule were observed, lower transition lines were adopted. 
{We determine molecular abundances relative to the H$_2$ column densities obtained in the previous study \cite{Pazukhin23}. All maps are smoothed to the same spatial resolution of $40\arcsec$, corresponding to the kinetic temperature and H$_2$ column density maps. Additionally, the H$_2$ column densities are obtained for the entire line of sight towards the clump, without the subtraction of the brightness value of the parent branch.
The maximum abundance relative to H$_2$ is obtained for HCN ($\sim$10$^{-8}$), and decreased to approximately 10$^{-10}$ for SiO. We note that abundance estimates give systematic errors due to LTE and optically thin assumptions, as well as the H$_2$ column densities derived from dust continuum emission. Thus, systematic errors can be of an order of magnitude for molecular abundances. 

In~\cite{Vasyunina11} a survey at 3 mm towards 15 IRDCs with the Mopra-22m telescope was conducted and molecular abundances of $\rm N_2H^+$, HCO$^+$, HNC, SiO, C$_2$H, HC$_3$N, and HNCO were obtained. These abundances are in good agreement with those determined in~\cite{Sanhueza12} for a similar sample of sources. However, the relative abundances obtained in our study are approximately one order of magnitude higher. In~\cite{Sanhueza12} it was also found that the abundances of $\rm N_2H^+$ and HCO$^{+}$ increase with chemical evolution.

In~\cite{YuXu16} 87 RMS sources (28 massive YSOs and 59 \Hii\ regions) are investigated and abundances of $\rm N_2H^+$, C$_2$H, HC$_3$N, and HNC are calculated. Their abundances were relatively low compared to IRDCs. In~\cite{Gerner14} a sample of 59 sources is observed and the chemical evolution is modelled including different evolutionary stages. The abundances is increased with evolutionary phase. More complex and heavy molecules are formed with evolving age until the HMC phase and decline for the UC\Hii\ stage,  when they are destroyed by the UV-radiation from the embedded sources. Their molecular abundances are in agreement with our results, assuming uncertainties of about 10.

\section{Conclusions}\label{sum}

In this work, we have carried out a multi-line survey at 2 and 3--4~mm towards 5 regions of massive star and stellar cluster formation, in order to study the physical and chemical properties at different evolutionary stages. We observed HCN, HNC, HCO$^+$, HC$_3$N, HNCO, OCS, CS, SiO, SO$_2$, and SO lines with the IRAM-30m radio telescope. The 850\micron\ dust emission data from the SCUBA Legacy catalogue is used to identify clumps. The RMS catalogue is used to classify them. We investigate the physical properties of the clumps and their molecular abundances. The results are as follows:
\begin{enumerate}

\item A total of 20 clumps were identified. Three clumps were found to be associated with the \Hii\ regions, 10 with YSOs, and 7 with submillimetre emission. The clumps have sizes of typical about 0.2~pc and masses ranging from 1 to 100\msun, kinetic temperatures ranging from 20 to 40~K and line widths of $\rm H^{13}CO^{+}$ (1–0) approximately 2\kms. 

\item We found no significant correlation in the line width--size and the line width--mass relationships. However, a strong correlation is observed in the mass--size relationships. The virial analysis indicated that three clumps are gravitationally bound. The virial parameter depends on mass as $\alpha_{\rm vir}\propto M^{-0.65}$. Furthermore, we suggested that magnetic fields of about 1~mG provide additional support for clump stability. 

\item The detection rates of molecular lines in clumps with YSO and \Hii\ regions are higher than those in submm.
The average line widths are about 3\kms, with a minimum for C$^{34}$S (2.4\kms) and a maximum for SiO (5.5\kms). In general, line widths increase with evolutionary stages. The maximum abundance relative to $H_2$ is obtained for HCN ($\sim$10$^{-8}$), and decreased to approximately 10$^{-10}$ for SiO.
\end{enumerate}

\begin{acknowledgements}
The authors are grateful to the anonymous reviewer for constructive comments that improved the paper quality.
This study was supported by the Russian Science Foundation grant No. 24-12-00153.
The research is based on observations made by the 041-19 project with the 30-m telescope. IRAM is supported by INSU/CNRS (France), MPG (Germany), and IGN (Spain). This research made use of the SIMBAD database and the VizieR catalogue access tool operated at CDS, Strasbourg, France \cite{2000A&AS..143....9W,10.26093/cds/vizier}. The original description 
of the VizieR service was published in~\cite{vizier2000}. This paper also made use of information from the RMS survey data base at \url{http://rms.leeds.ac.uk/} which was constructed with support from the Science and Technology Facilities Council of the UK.  This research made use of \texttt{NumPy} \cite{numpy}, \texttt{LMFIT} \cite{lmfit}, \texttt{astropy} \cite{astropy}, \texttt{matplotlib} \cite{matplotlib}, and \texttt{SciPy} \cite{scipy}. This research made use of astrodendro, a Python package to compute dendrograms of Astronomical data (\url{http://www.dendrograms.org}).
\end{acknowledgements}

%
%
\bibliographystyle{maik}
\bibliography{main}

\newpage

\begin{table} 
\setcaptionmargin{0mm}
\onelinecaptionstrue
\captionstyle{flushleft} 
\caption{List of sources} \label{tab:source}
\bigskip
\begin{tabular}{l|c|c|c|c|c}
 \hline
            Source & RA(J2000) & Dec(J2000) & $V_{\rm lsr}$ & $d^{(a)}$ & Note \\
                   & (h:m:s) & (d:m:s) & ($\rm km\,s^{-1}$) & (kpc) & \\
            \hline
            L1287     & 00:36:47.5 & 63:29:02.1 & $-17.7$ & $0.93\pm0.03$ & G121.30+0.66, IRAS~00338+6312\\
            S187     & 01:23:15.4 & 61:49:43.1 & $-14.0$ & $1.44\pm0.26$   & G126.68–0.81, IRAS~01202+6133 \\
            S231     & 05:39:12.9 & 35:45:54.0 & $-16.6$ & $1.56\pm0.09$  & G173.48+2.45, IRAS~05358+3543 \\
            DR~21(OH) & 20:39:00.6 & 42:22:48.9 &  $-3.8$ & $1.50\pm0.08$  & G81.72+0.57 \\
            NGC~7538  & 23:13:44.7 & 61:28:09.7 & $-57.6$ & $2.65\pm0.12$  & G111.54+0.78, IRAS~23116+6111 \\
            \hline
\end{tabular}       
\begin{flushleft}
   (a) Distances to sources are quoted from~\cite{Rygl10,Russeil07,Burns15,Rygl12,Moscadelli09}.
    
\end{flushleft}
\end{table}

\begin{table} 
\setcaptionmargin{0mm}
\onelinecaptionstrue
\captionstyle{flushleft} 
\caption{List of molecular lines} \label{tab:lines}
\bigskip
\begin{tabular}{l|c|c|c}
        \hline
        Molecule & Transition & Rest frequency & $E_{u}/k_{\rm B}$ \\
         &  & (MHz) & (K) \\
        \hline
	SO$_2$& $ 6(0,6) - 5(1,5)$ & 72758.243 & 19.2  \\
 
        HC$_3$N	&	$8-7$& 72783.822    & 15.7\\	
                 & $10-9$ & 90979.023   & 24.0	\\
                 & $17-16$ & 154657.284 & 66.8\\	
        OCS     & $6-5$ &  72976.7794 & 12.3 \\
                & $7-6$ &  85139.1032 & 16.3 \\        
        SO & $2-1$	& 86093.950 & 19.3 \\
           & $4-3$  &	138178.600 & 15.9\\
        H$^{13}$CN & $ 1-0$ & 86339.921 & 4.1 \\
        H$^{13}$CO$^{+}$ & $1-0$ & 86754.288  & 4.2 \\
        SiO & $2-1$ & 86846.985  & 6.3 \\
        HN$^{13}$C & $ 1-0$ & 87090.825 & 4.2 \\
        HNCO& $4(0,4)-3(0,3)$ & 87925.237 & 10.5 \\
        HCN & $ 1-0$ & 88631.602 & 4.3 \\
        HCO$^{+}$ & $ 1-0$ & 89188.525 & 4.3 \\
        HNC & $ 1-0$ & 90663.568 & 4.4 \\
        C$^{34}$S & $ 3-2$ & 144617.101 & 13.9 \\

            \hline     
\end{tabular}
\end{table}

\begin{table}
\scriptsize
\setcaptionmargin{0mm}
\onelinecaptionstrue
\captionstyle{flushleft} 
\caption{List of indentified clumps} \label{tab:clumps}
\bigskip
\begin{tabular}{l|c|c|c|c|c|c c|c|c|c|c}
 
\hline
Source & id  & RA(J2000)   &   Dec(J2000)  &    $F_{\rm \nu}^{(a)}$ & $F_{\rm bg}$ & \multicolumn{2}{c|}{$R_{\rm eff}$}  & $L_{\rm bol}^{(b)}$ & Masers$^{(c)}$ & Outflows & Type \\ 
&  & (h:m:s)   &   (d:m:s)   & (Jy) & (Jy/beam) & (pc) & ($^{\prime\prime}$) & ($10^{3}$$\times$$ L_{\odot}$) &  &  &  \\ 
\hline
L1287
&  1 &   00:36:58.4 & 63:27:59.3 &   0.20 (0.01) & 0.45 (0.04) &   0.07 & 17 & - & - & - & -\\
&  2 &   00:37:03.9 & 63:28:03.0 &   0.07 (0.01) & 0.45 (0.04) &   0.05 & 11& - & - & - & -\\
&  3 &   00:36:47.4 & 63:28:59.6 &  12.20 (0.03) & 0.41 (0.04) &  0.21 & 48 & 1 & H$_2$O, OH, MI, MII & HCO$^{+}$, SiO & YSO\\

S187 
 &  1 &   01:23:32.2 & 61:48:41.0 &   2.79 (0.04) & 0.89 (0.07) &  0.23 & 33& 2.6 & OH & - & YSO\\
&  2 &   01:23:13.2 & 61:50:01.7 &   0.85 (0.03) & 0.28 (0.07) &  0.19 & 28& - & - & HCO$^{+}$ & YSO$^{ (d)}$\\

S231
&  1  &  05:39:10.8 & 35:45:14.9 &   0.02 (0.00) &2.48 (0.04) &   0.05 & 7& 2.9 & H$_2$O, MI, MII & - & YSO\\
 &  2 &   05:39:12.3 & 35:45:49.8 &  5.29 (0.02) &2.54 (0.04) &    0.18 & 24& 7.6 & H$_2$O, OH, MI, MII & HCO$^{+}$, SiO & YSO\\
&  3 &   05:39:08.6 & 35:46:41.0 &   0.01 (0.00) &1.74 (0.04) &   0.04 & 6& - & - & - & -\\

DR~21(OH)
 &  1 &  20:39:01.0 & 42:22:09.3 &   0.34 (0.01) &14.54 (0.07) &   0.06 & 8& 1.9 & - & - & \Hii\ region\\
&  2  & 20:39:00.5 & 42:22:44.2 &  11.62 (0.02) &14.68 (0.07) &   0.14 & 19& 12 & H$_2$O, OH, MI, MII & HCO$^{+}$, SiO & \Hii\ region\\
&  3 &  20:38:59.5 & 42:23:29.1 &   0.05 (0.01) &7.73 (0.07) &   0.06 & 8& - & - & - & -\\
&  4 &  20:39:01.7 & 42:24:55.7 &  3.25 (0.03) &5.07 (0.07) &   0.16 & 21& 9 & H$_2$O, OH, MI, MII & HCO$^{+}$, SiO & YSO\\
&  5 & 20:39:03.1 & 42:25:46.8 &   0.92 (0.02) &5.02 (0.07) &   0.10 & 13& 2.6 &  MI, MII & HCO$^{+}$, SiO & YSO\\
&  6 & 20:39:00.1 & 42:27:32.4 &   0.10 (0.02) &0.60 (0.07) &   0.09 & 12& - & - & - & -\\
&  7  & 20:38:51.6 & 42:27:16.7 &   0.15 (0.02) &0.46 (0.07) &   0.11 & 16& - & - & - & -\\

NGC~7538
&  1 &  23:13:33.5 & 61:25:47.7 &   0.08 (0.02) &0.90 (0.08) &   0.14 & 11& - & - & - & -\\
&  2  & 23:13:45.0 & 61:26:50.7 &  13.25 (0.04) &3.88 (0.08) &   0.32 & 25& 4.6 &   H$_2$O, OH, MI & HCO$^{+}$, SiO & YSO\\
&  3 & 23:14:02.6 & 61:27:10.9 &  28.72 (0.05) &3.84 (0.08) &   0.38 & 30& 23 &  H$_2$O, MI & HCO$^{+}$ & YSO\\
&  4  & 23:13:44.8 & 61:28:03.7 &  5.51 (0.04) &1.18 (0.08) &   0.42 & 33& 210 &  H$_2$O, OH, MI, MII & HCO$^{+}$, SiO & \Hii\ region\\
&  5  & 23:13:29.7 & 61:29:01.1 &   2.64 (0.04) &0.92 (0.08) &   0.39 & 31& 5.6 &  H$_2$O & - & YSO\\
\hline
\end{tabular}     
\begin{flushleft}
      (a) Integrated flux of a leaf/clump (with parent branch brightness ($F_{\rm bg}$) subtracted). \\
      (b)  The data on luminosity and associated source type are taken from the RMS source catalogue~\citep{Lumsden13}. \\ (b) The MI and MII indicate Class I and Class II methanol masers, respectively. \\ (c) pre-main-sequence star S187H$\alpha$~\citep{Zavagno94}. \\
     The errors for each parameter are shown in parentheses.    
\end{flushleft}
\end{table}

\begin{table}

\scriptsize
\setcaptionmargin{0mm}
\onelinecaptionstrue
\captionstyle{flushleft} 
\caption{List of clump parameters} \label{tab:params}
\bigskip
\begin{tabular}{l|c|c|c|c|c|c|c|c|c}
 \hline

Source & id  & $\sigma_{\rm tot}$ & $T_{\rm kin}$ & $N (\rm H_{2})$ & $\mathcal{M}$  & $M$ & $M_{cor}^{(a)}$ & $M_{\rm vir}$ & $\alpha_{\rm vir}$  \\ 
& &   ($\rm km\,s^{-1}$) & (K) & ($10^{22}$\pcm) &   & ($M_{\odot}$)& ($M_{\odot}$)& ($M_{\odot}$) &  \\ 
\hline
\multicolumn{10}{c}{YSO}   \\
\hline
L1287
&  3 & 0.85 (0.04) & 20.61 (0.73) & 0.25 (0.02) & 3.00 (0.14) & 79.8 (3.6) & 56.9 (2.6) & 178.6 (7.2) & 3.1 (0.1) \\
S187
&  1 & 0.69 (0.14) & 22.10 (1.17) & 0.40 (0.14) & 2.29 (0.48) & 89.6 (22.9) & 31.2 (8.0) & 129.8 (25.3) & 4.2 (0.9) \\
&  2 & 0.50 (0.05) & 18.02 (0.66) & 0.16 (0.06) & 1.73 (0.17) & 22.1 (5.7) & 9.6 (2.5) & 55.7 (8.1) & 5.8 (1.2) \\
S231
&  1 & 0.98 (0.10) & 29.99 (1.57) & 0.78 (0.29) & 2.85 (0.29) & 8.1 (0.7) & 0.3 (0.1) & 56.9 (4.6) & 226.7 (44.6) \\
&  2 & 1.34 (0.10) & 24.95 (0.98) & 1.25 (0.14) & 4.45 (0.33) & 174.3 (14.2) & 69.4 (5.7) & 388.3 (25.7) & 5.6 (0.4) \\
DR~21(OH)
&  4 & 0.94 (0.02) & 22.91 (0.44) & 1.88 (0.20) & 3.18 (0.07) & 188.3 (14.2) & 39.4 (3.0) & 160.4 (6.5) & 4.1 (0.2) \\
&  5 & 0.97 (0.02) & 21.80 (0.46) & 1.87 (0.21) & 3.38 (0.07) & 66.5 (5.0) & 11.2 (0.9) & 104.4 (4.2) & 9.4 (0.6) \\
NGC~7538
&  2 & 1.81 (0.05) & 28.79 (0.61) & 2.73 (0.25) & 5.63 (0.14) & 990.5 (63.4) & 501.6 (32.2) & 1228.0 (45.1) & 2.4 (0.1) \\
&  3 & 1.58 (0.35) & 29.55 (1.50) & 0.73 (0.07) & 4.82 (1.07) & 1921.8 (123.1) & 1087.0 (69.6) & 1104.0 (177.5) & 1.0 (0.1) \\
&  5 & 0.83 (0.19) & 26.50 (0.67) & 0.38 (0.04) & 2.54 (0.58) & 270.9 (17.4) & 99.8 (6.6) & 315.9 (51.6) & 3.2 (0.4) \\

\hline
\multicolumn{10}{c}{\Hii\ region}  \\
\hline
DR~21(OH)
&  1 & 1.39 (0.02) & 23.50 (0.26) & 5.31 (0.61) & 4.75 (0.05) & 68.2 (5.1) & 4.1 (0.3) & 135.4 (5.2) & 33.2 (2.1) \\
&  2 & 1.21 (0.04) & 24.29 (0.72) & 5.96 (0.64) & 4.04 (0.14) & 486.3 (36.7) & 140.9 (10.6) & 237.0 (10.6) & 1.7 (0.1) \\
NGC~7538
&  4 & 1.53 (0.07) & 39.80 (1.34) & 2.57 (0.23) & 3.98 (0.18) & 417.8 (26.8) & 208.6 (13.5) & 1153.0 (52.5) & 5.5 (0.3) \\

\hline
\multicolumn{10}{c}{submm}  \\
\hline
L1287
&  1 & 0.74 (0.03) & 18.07 (0.69) & 0.19 (0.02) & 2.76 (0.12) & 4.0 (0.2) & 0.9 (0.1) & 47.1 (1.8) & 51.3 (3.1) \\
&  2 & 0.59 (0.06) & 16.73 (1.03) & 0.16 (0.03) & 2.23 (0.22) & 1.7 (0.1) & 0.3 (0.0) & 20.6 (1.5) & 63.7 (6.6) \\
S231
&  3 & 1.27 (0.10) & 24.96 (1.64) & 0.50 (0.31) & 4.19 (0.32) & 4.3 (0.4) & 0.1 (0.1) & 83.1 (5.7) & 646.7 (201.8) \\
DR~21(OH)
&  3 & 1.40 (0.03) & 22.74 (0.61) & 2.89 (0.90) & 4.89 (0.10) & 29.0 (2.2) & 0.6 (0.1) & 126.2 (5.1) & 227.2 (36.0) \\
&  6 & 0.68 (0.04) & 18.37 (0.72) & 0.20 (0.05) & 2.49 (0.15) & 7.0 (0.6) & 1.2 (0.2) & 47.5 (2.8) & 38.4 (4.8) \\
&  7 & 0.83 (0.11) & 21.09 (0.65) & 0.15 (0.03) & 2.89 (0.38) & 8.9 (0.7) & 1.8 (0.3) & 90.2 (9.0) & 49.8 (6.3) \\
NGC~7538
&  1 & 0.97 (0.06) & 23.51 (1.06) & 0.27 (0.08) & 3.22 (0.19) & 23.8 (1.6) & 3.1 (0.6) & 148.9 (7.7) & 47.8 (6.9) \\

\hline
\end{tabular}     
\begin{flushleft}
      (a) Corrected values obtained after subtraction of parent branch flux. \\ The errors for each parameter are shown in parentheses.
\end{flushleft} 
\end{table}

\begin{figure}
\setcaptionmargin{5mm}
\onelinecaptionsfalse 

\captionstyle{normal}

    \begin{minipage}{0.33\linewidth}
        \centering \includegraphics[width=\linewidth]{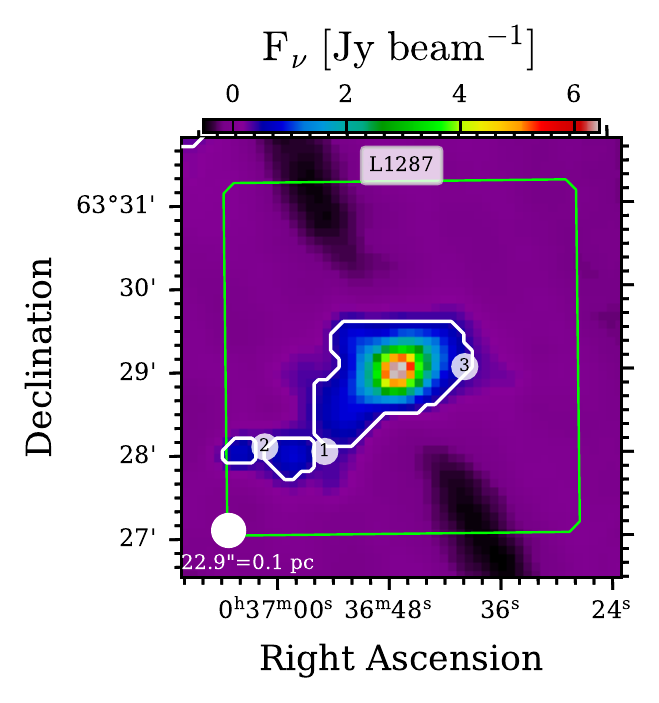}
    \end{minipage}\hfill
    \begin{minipage}{0.33\linewidth}
        \centering \includegraphics[width=\linewidth]{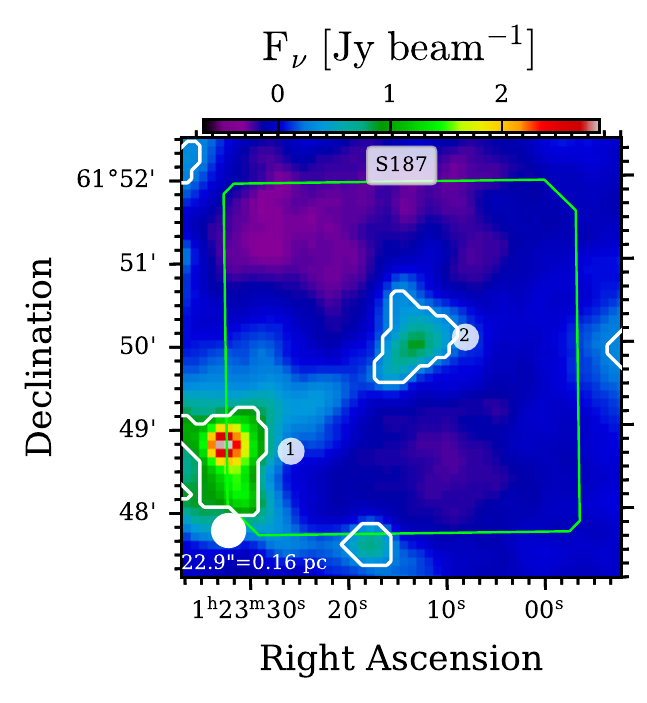}
    \end{minipage}\hfill
    \begin{minipage}{0.33\linewidth}
        \centering \includegraphics[width=\linewidth]{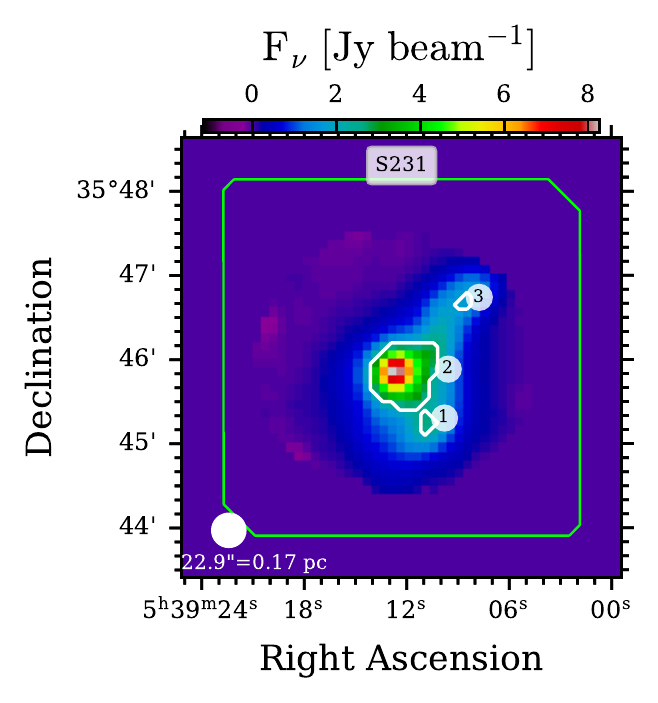}
    \end{minipage}   
    \begin{minipage}{0.45\linewidth}
        \centering \includegraphics[width=\linewidth]{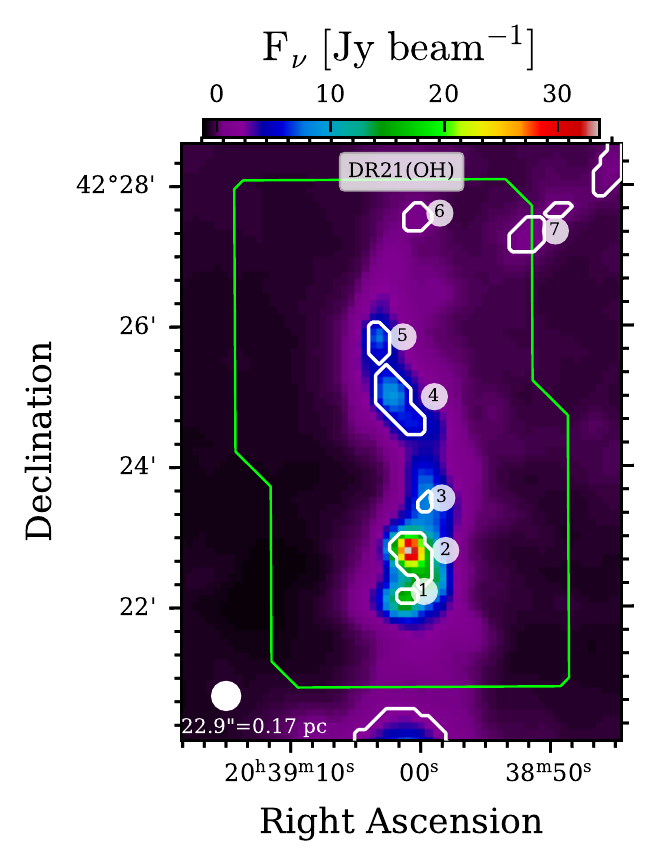}
    \end{minipage}\hfill
    \begin{minipage}{0.45\linewidth}
        \centering \includegraphics[width=\linewidth]{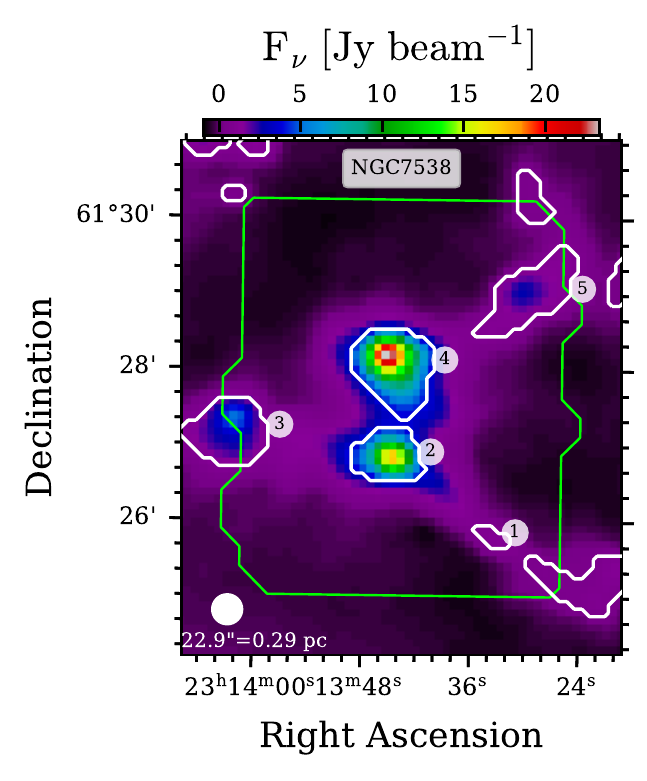}
    \end{minipage}\hfill
   \caption{Maps of the dust emission according to the SCUBA Legacy catalogue at 850\micron. The clumps obtained using the astrodenro method are highlighted with contours and numbers. The source is indicated in the upper center of each panel. The green outline demonstrates the IRAM-30m map region. The beam size and the scale equivalent are shown in the lower left corner of each panel.} \label{fig:clump}
\end{figure}

\begin{figure}
\setcaptionmargin{5mm}
\onelinecaptionsfalse 

\captionstyle{normal}
\centering \includegraphics[width=\linewidth]{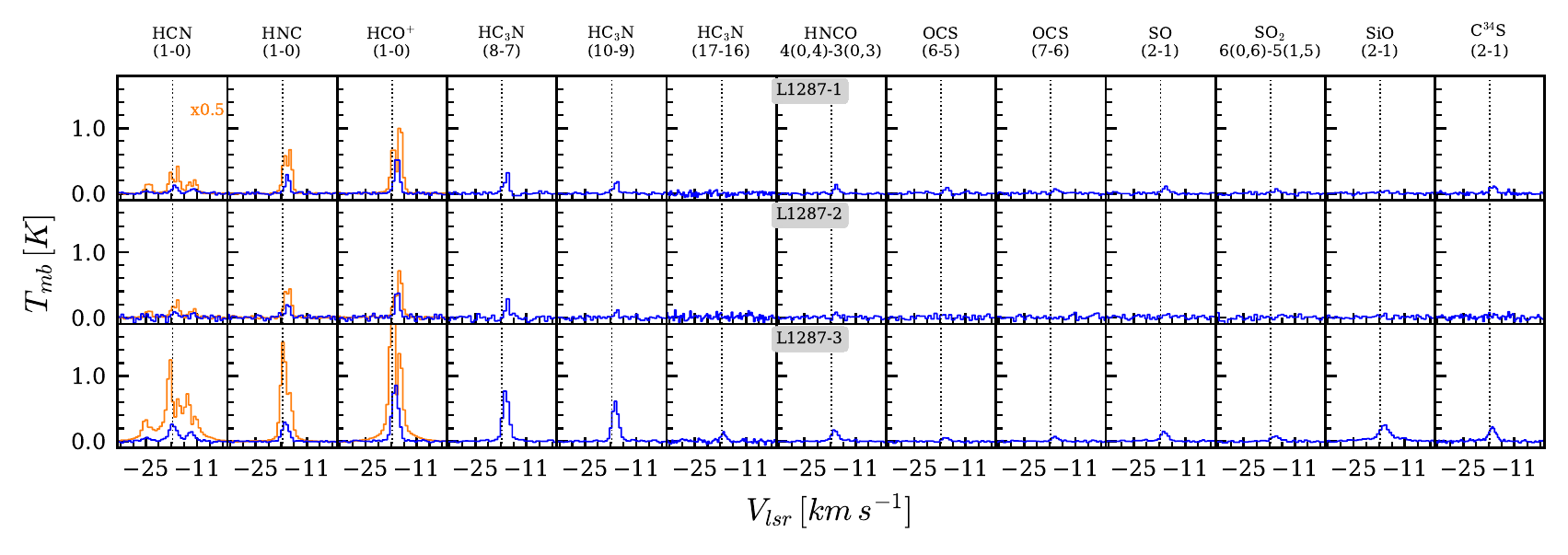}
   \caption{Spectra are extracted towards clumps of L1287. {Spectra for $^{13}$C isotopologues (blue colour) of HCN, HNC, and HCO$^{+}$ and halved spectra for main isotopologues (orange colour) are shown in the first three columns.} The system velocity is represented by a grey dashed line on each panel. The transitions are indicated at the top of each column. The clump identifiers are listed in the centre. The spectra of the other sources are presented in the Appendix (Fig.~\ref{fig:add_spectra})} \label{fig:spectra}
\end{figure}

\begin{figure}
\setcaptionmargin{5mm}
\onelinecaptionsfalse 

\captionstyle{normal}
        \centering \includegraphics[width=.9\linewidth]{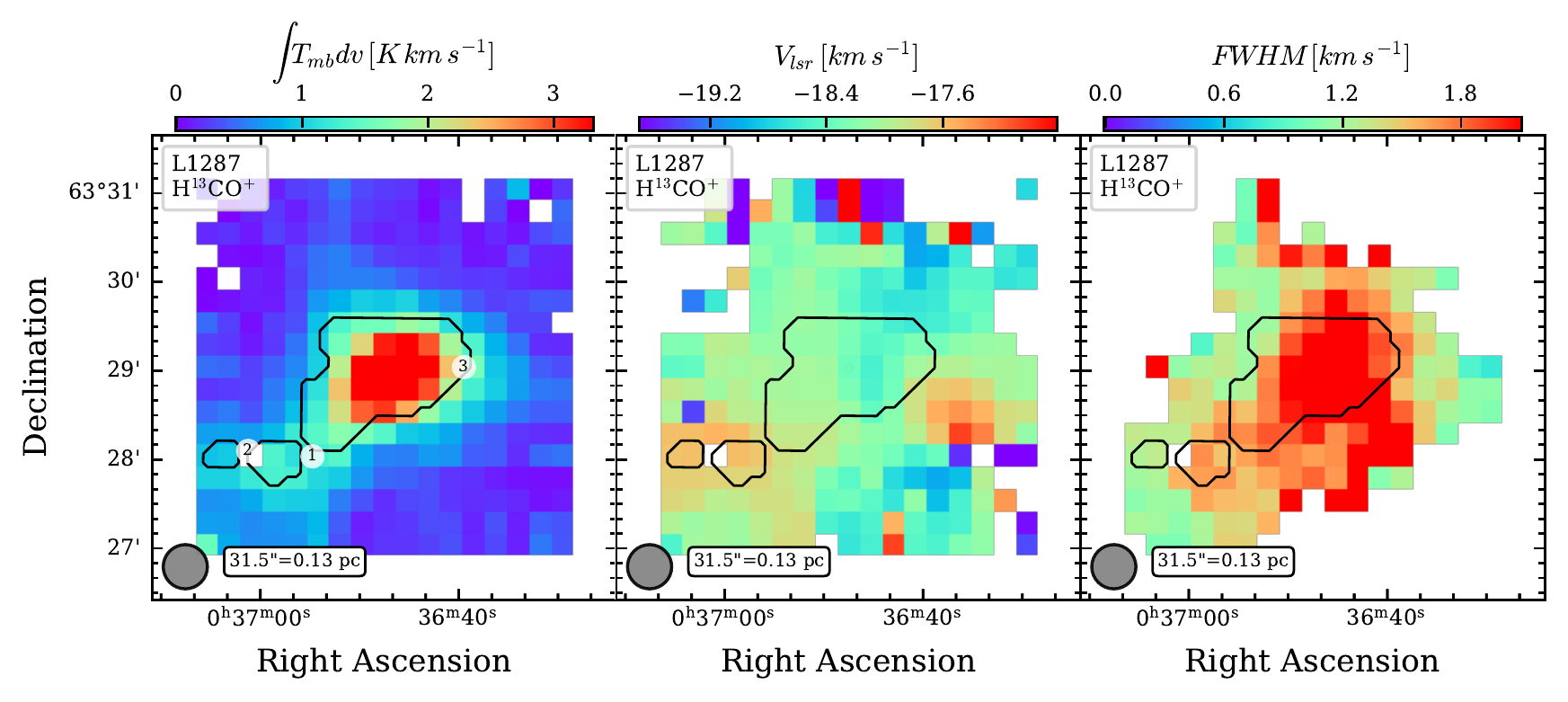} 
   \caption{Maps of the integrated intensity, velocity and line width of the line J=1-0 $\rm H^{13}CO^{+}$ for L1287. The contours illustrate the derived clumps. The clump ids are the same as in Fig.\ref{fig:clump}. The source is indicated in the upper left corner of each panel. The beam size and the scale equivalent are shown in the lower left corner of each panel. The maps of the other sources are presented in the Appendix (Fig.~\ref{fig:add_moment})} \label{fig:moment}
\end{figure}

\begin{figure}
\setcaptionmargin{5mm}
\onelinecaptionsfalse 

\captionstyle{normal}
\centering \includegraphics[width=.5\linewidth]{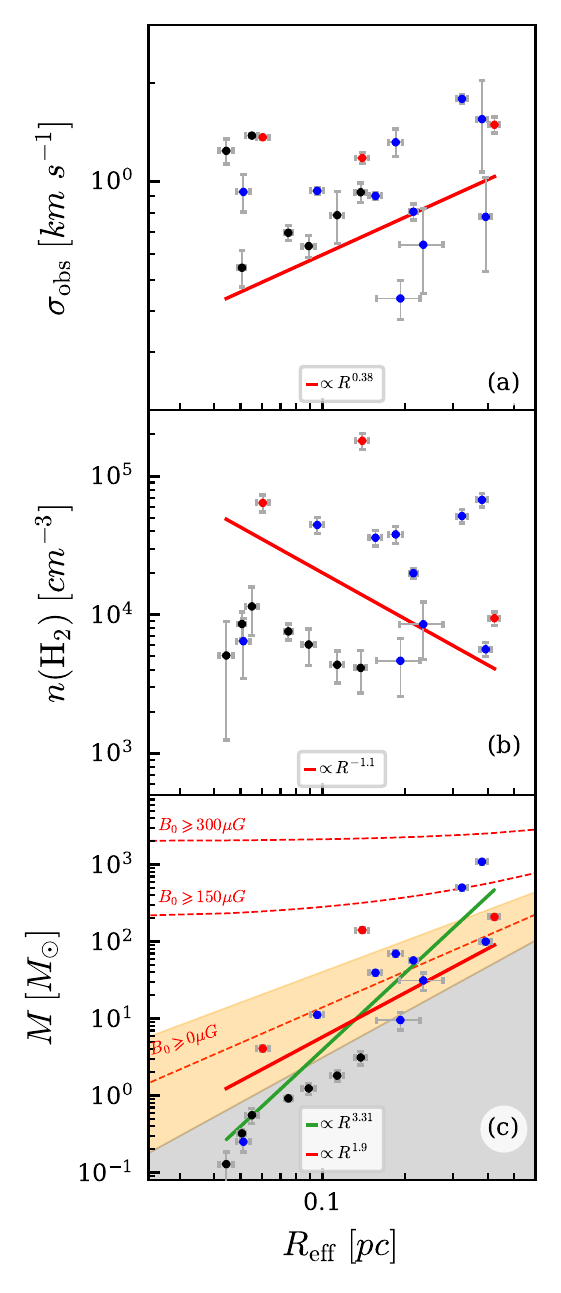}
   \caption{Relation of (a) $\sigma_{\rm obs}-R_{\rm eff}$, (b) $n({\rm H_2})-R_{\rm eff}$ and (c) $M-R_{\rm eff}$. The submm, YSO and \Hii\ region clumps are shown in black, blue and red, respectively. The green line shows the fitting result. The line slopes are presented in each panel. The red line demonstrates the original Larson relation~\cite{Larson81}. The designation in panel (c) was adapted from \cite{Pillai15}. The minimum "critical"\ magnetic field strengths are indicated by the red dashed lines. Diffuse clouds, indicated by the grey shading, and low-mass star-forming regions without magnetic support, highlighted by orange shading. } \label{fig:size}
\end{figure}
\begin{figure}
\setcaptionmargin{5mm}
\onelinecaptionsfalse 

\captionstyle{normal}
\centering \includegraphics[width=.5\linewidth]{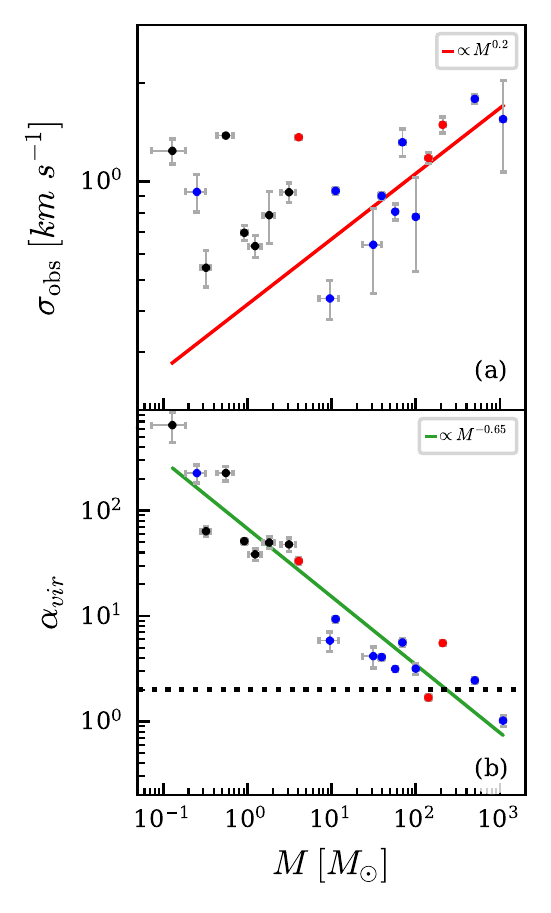}
   \caption{Relation of (a) $\sigma_{\rm obs}-M$ and (b) $\alpha_{\rm vir}-M$. The submm, YSO and \Hii\ region clumps are shown in black, blue and red, respectively. The green line indicates the fitting result. The line slopes are presented in each panel. The red line demonstrates the original Larson relation~\cite{Larson81}. In panel (b), the horizontal dashed line at $\alpha_{\rm vir} = 2$ gives the critical value expected for non-magnetised clouds.} \label{fig:mass}
\end{figure}

\begin{figure}
\setcaptionmargin{5mm}
\onelinecaptionsfalse 

\captionstyle{normal}

\centering \includegraphics[width=\linewidth]{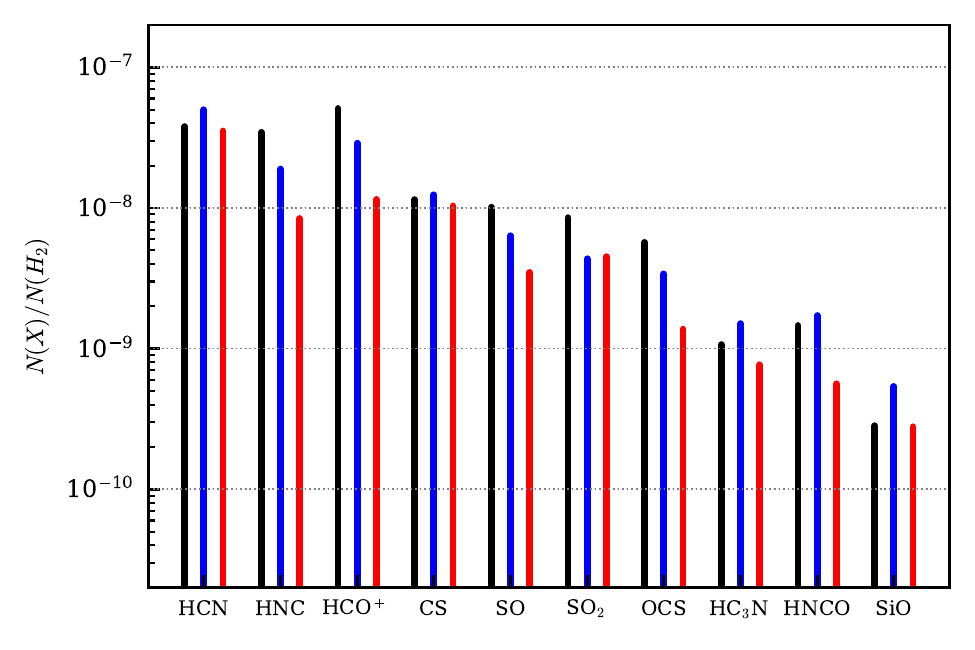}

   \caption{Averaged molecular abundances relative to H$_2$ for different clump types. The submm, YSO, and \Hii\ region clumps are shown in black, blue, and red, respectively.} \label{fig:abun_clump}
\end{figure}

\begin{appendix}
\FloatBarrier
\section{Spectra}
\FloatBarrier

\begin{figure}
\setcaptionmargin{5mm}
\onelinecaptionsfalse 

\captionstyle{normal}
\centering \includegraphics[width=\linewidth]{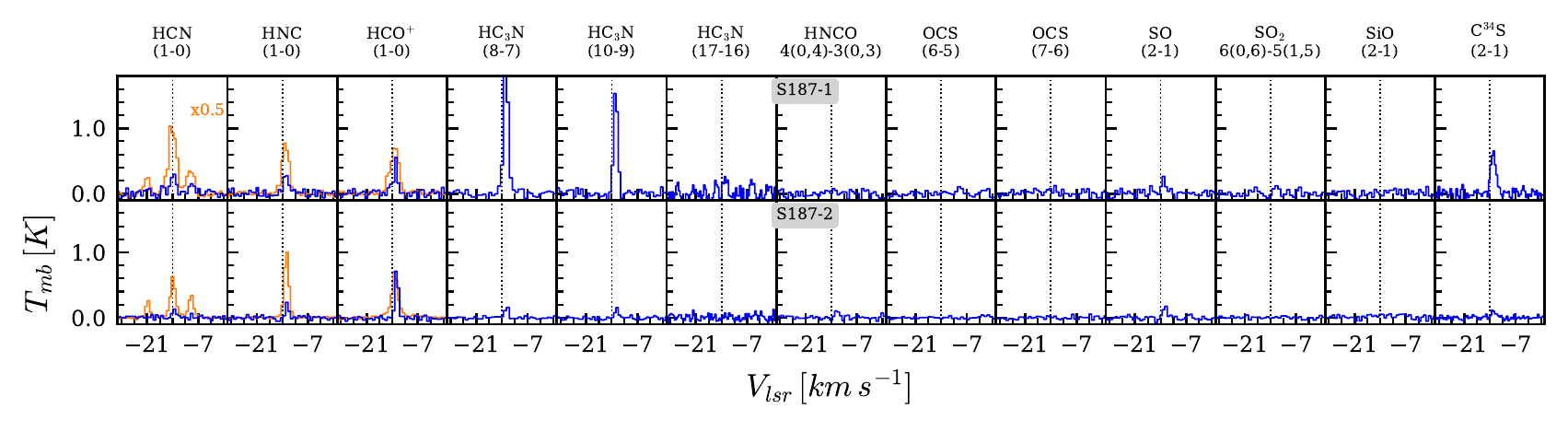}
   \caption{Spectra are extracted towards clumps of S187. {Spectra for $^{13}$C isotopologues (blue color) of HCN, HNC, and HCO$^{+}$ and halved spectra for main isotopologues (orange color) are shown in the first three columns.} The system velocity is represented by a grey dashed line on each panel. The transitions are indicated at the top of each column. The clump identifiers are listed in the centre.} \label{fig:add_spectra}
\end{figure}
\addtocounter{figure}{-1}
\begin{figure}
\setcaptionmargin{5mm}
\onelinecaptionsfalse 

\captionstyle{normal}
        \centering \includegraphics[width=\linewidth]{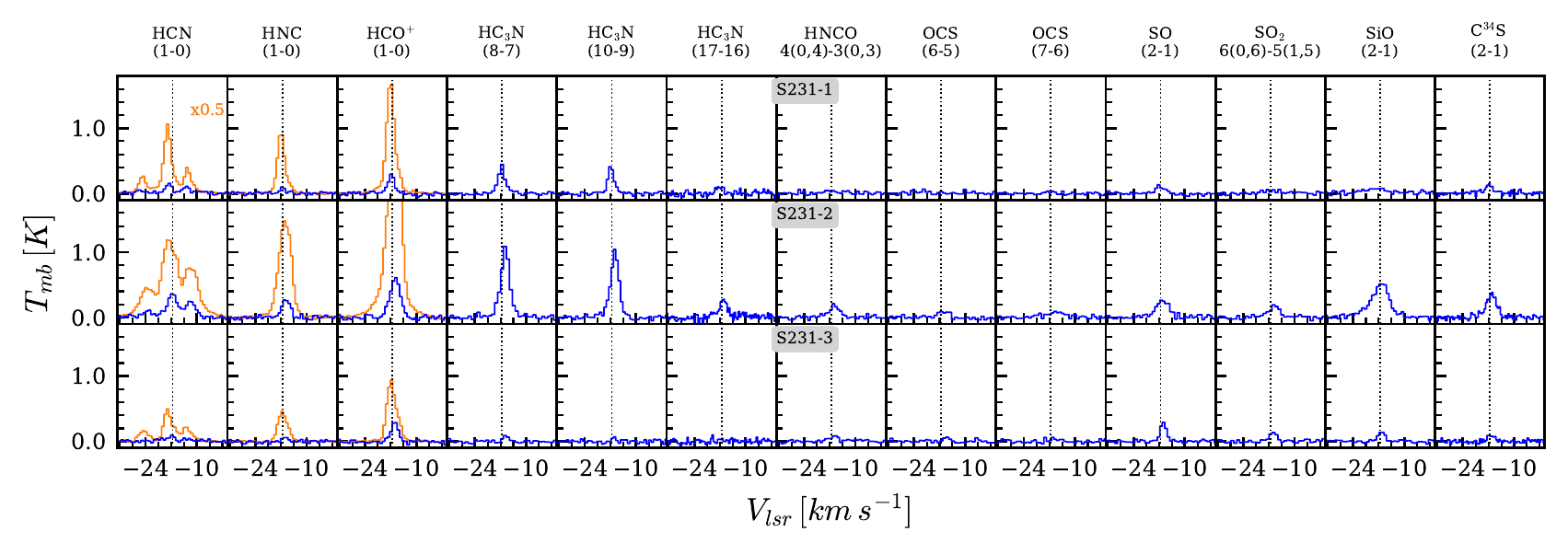}
   \caption{continued.}
\end{figure}
\addtocounter{figure}{-1}
\begin{figure}
\setcaptionmargin{5mm}
\onelinecaptionsfalse 

\captionstyle{normal}
        \centering \includegraphics[width=\linewidth]{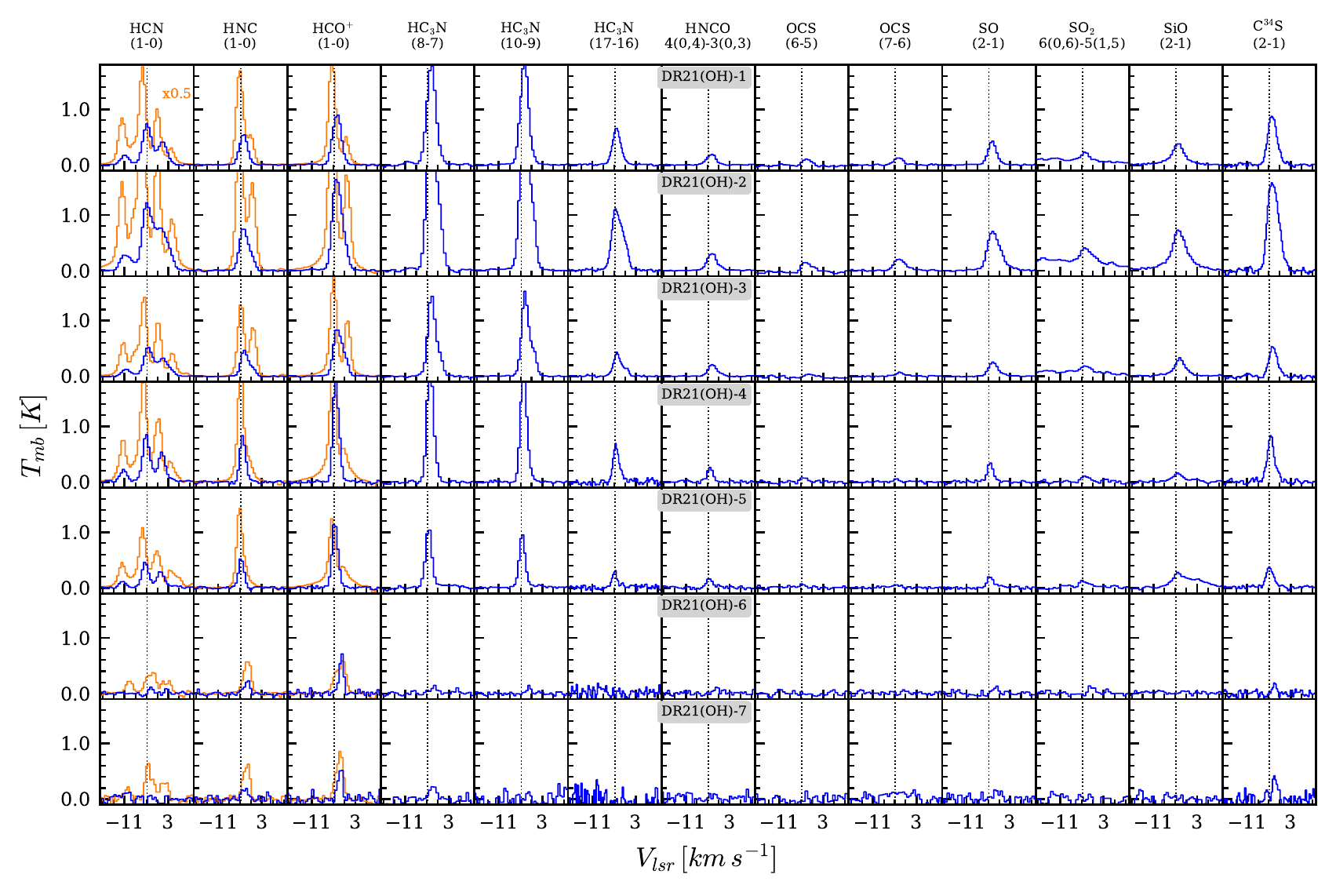}
   \caption{continued.}
\end{figure}
\addtocounter{figure}{-1}
\begin{figure}
        \centering \includegraphics[width=\linewidth]{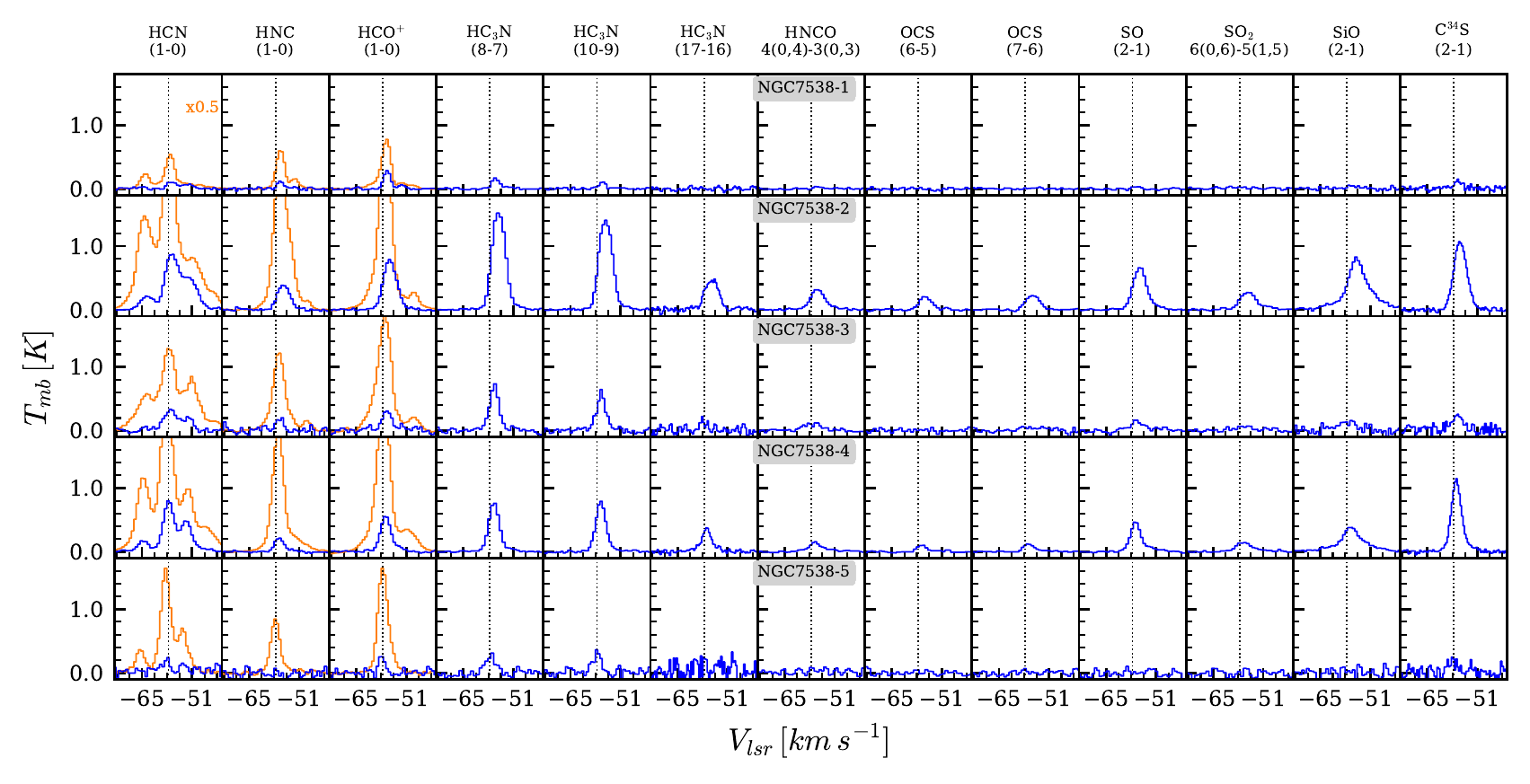}
   \caption{continued.}
\end{figure}

\FloatBarrier
\section{Moment maps}
\FloatBarrier
\begin{figure}
\setcaptionmargin{5mm}
\onelinecaptionsfalse 

\captionstyle{normal}
        \centering \includegraphics[width=.9\linewidth]{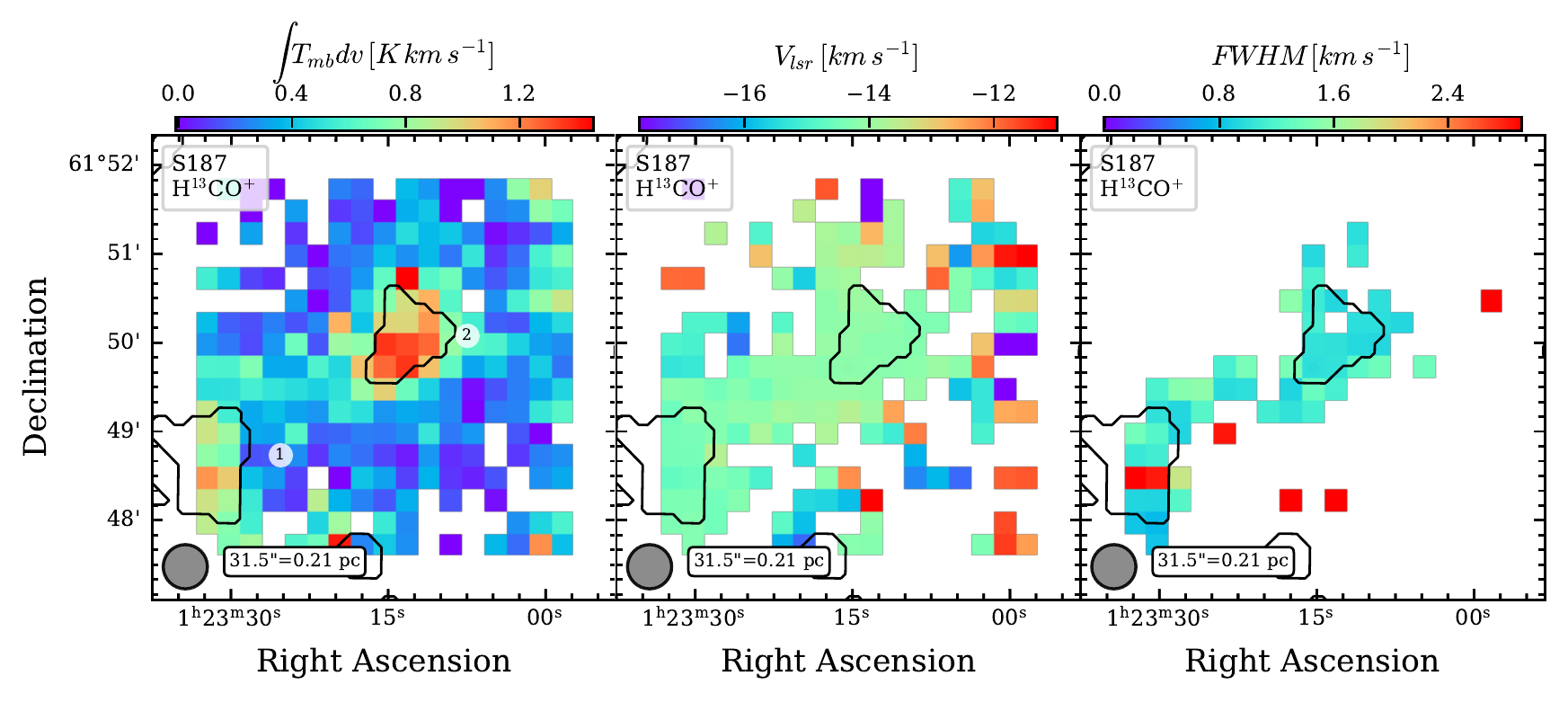}
   \caption{Maps of the integrated intensity, velocity and line width of the line J=1-0 $\rm H^{13}CO^{+}$ for S187. The contours illustrate the derived clumps. The clump ids are the same as in Fig.\ref{fig:clump}. The source is indicated in the upper left corner of each panel. The beam size and the scale equivalent are shown in the lower left corner of each panel.} \label{fig:add_moment}
\end{figure}
\addtocounter{figure}{-1}
\begin{figure}
\setcaptionmargin{5mm}
\onelinecaptionsfalse 

\captionstyle{normal}
        \centering \includegraphics[width=\linewidth]{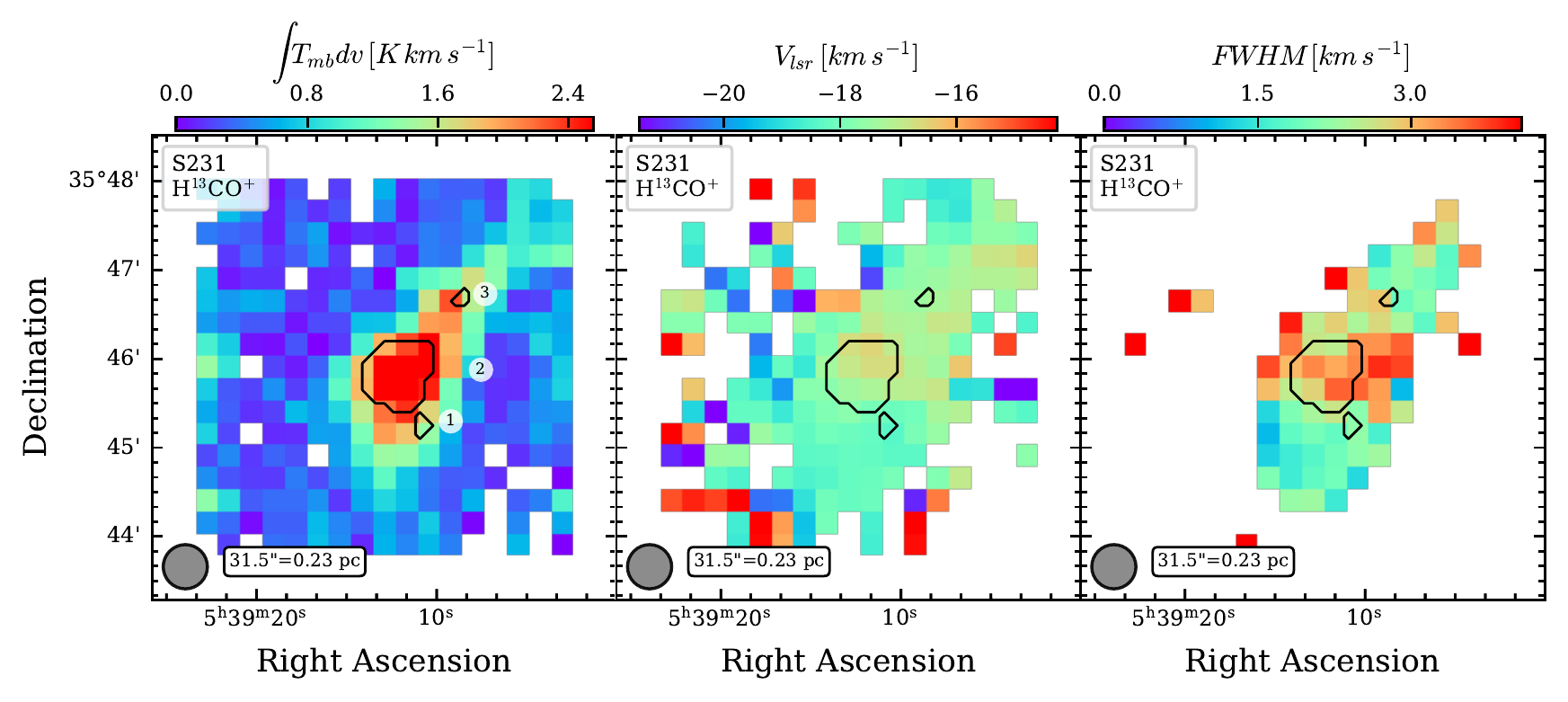}
   \caption{continued.}
\end{figure}
\addtocounter{figure}{-1}
\begin{figure}
\setcaptionmargin{5mm}
\onelinecaptionsfalse 

\captionstyle{normal}
        \centering \includegraphics[width=\linewidth]{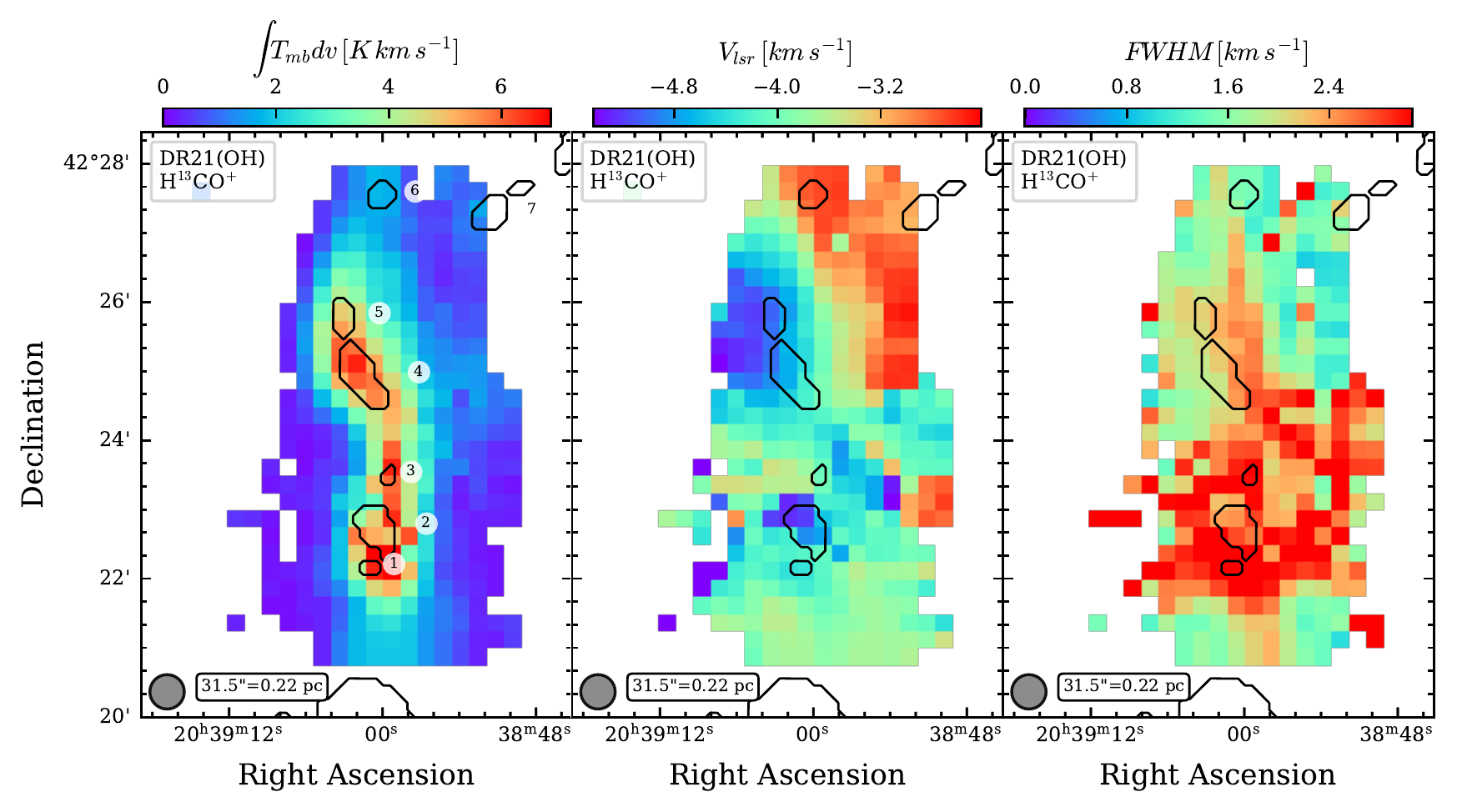}
   \caption{continued.}
\end{figure}
\addtocounter{figure}{-1}
\begin{figure}
\setcaptionmargin{5mm}
\onelinecaptionsfalse 

\captionstyle{normal}
        \centering \includegraphics[width=\linewidth]{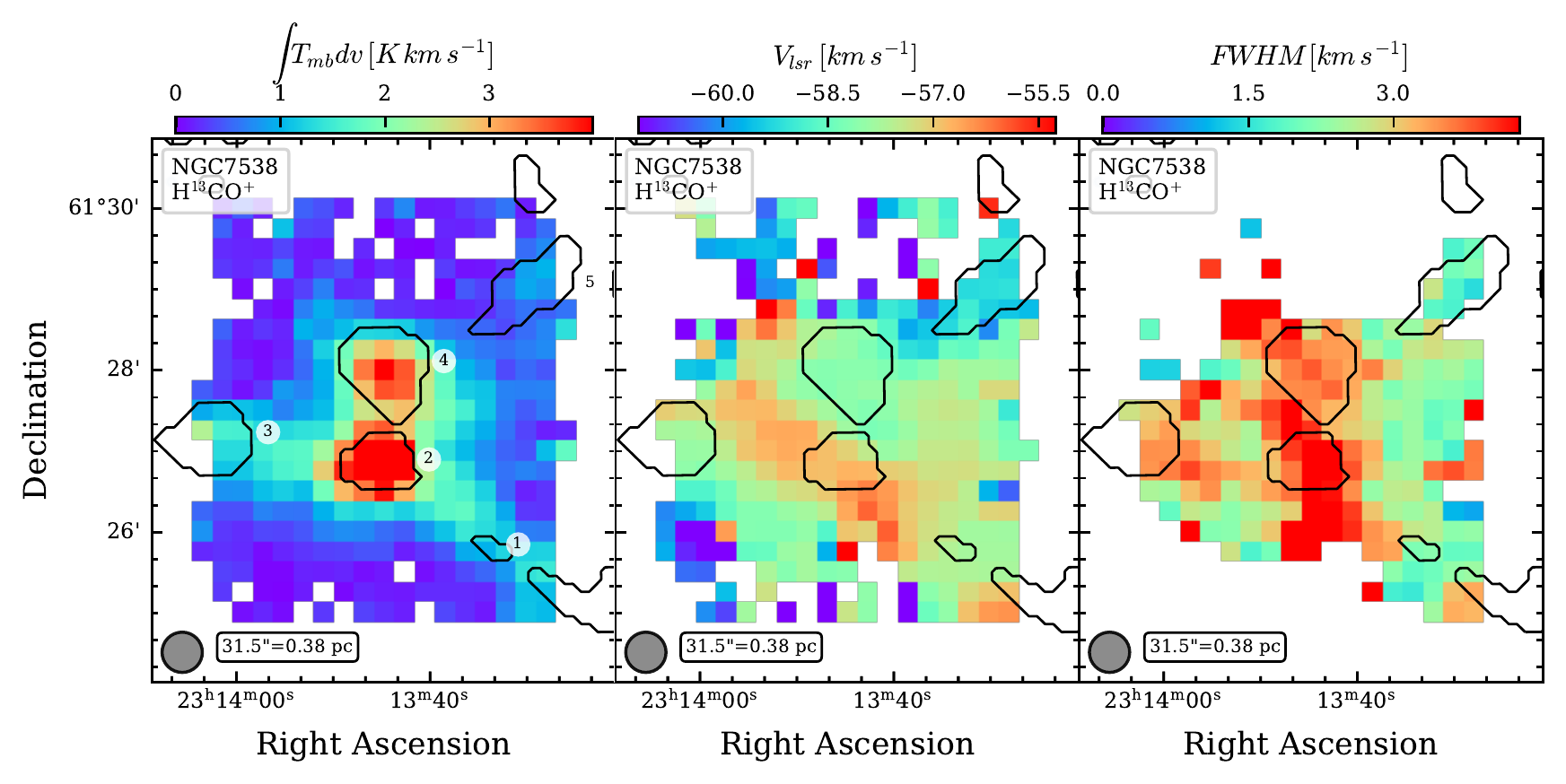}
   \caption{continued.}
\end{figure}

\FloatBarrier
\section{3-color WISE images}
\FloatBarrier
\begin{figure}
\setcaptionmargin{5mm}
\onelinecaptionsfalse 

\captionstyle{normal}
\begin{minipage}{0.33\linewidth}
        \centering \includegraphics[width=\linewidth]{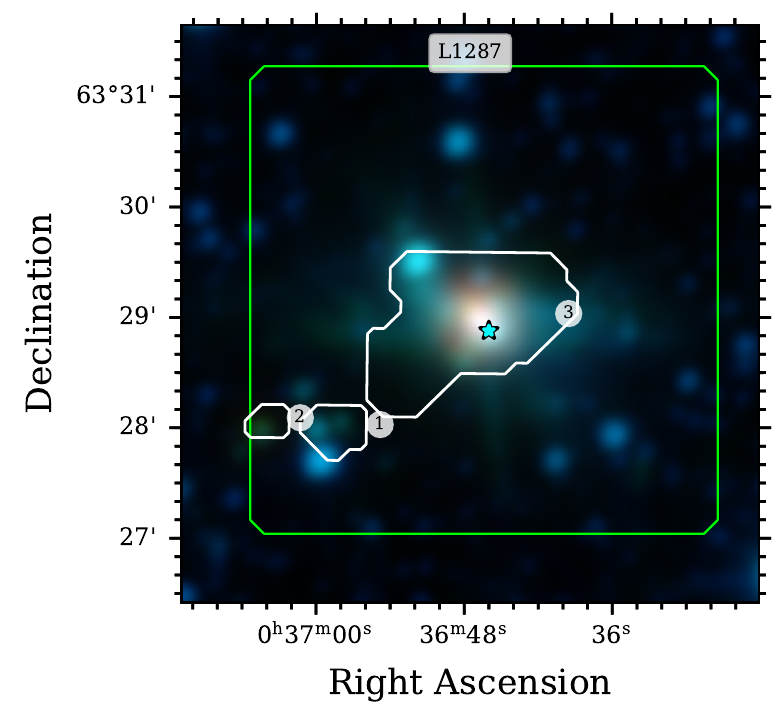}
    \end{minipage}\hfill
    \begin{minipage}{0.33\linewidth}
        \centering \includegraphics[width=\linewidth]{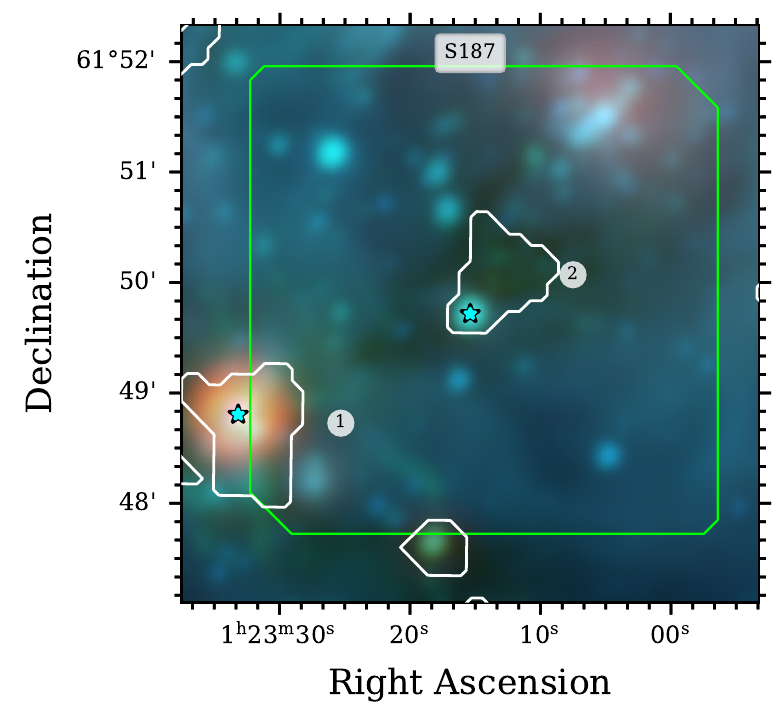}
    \end{minipage}\hfill
    \begin{minipage}{0.33\linewidth}
        \centering \includegraphics[width=\linewidth]{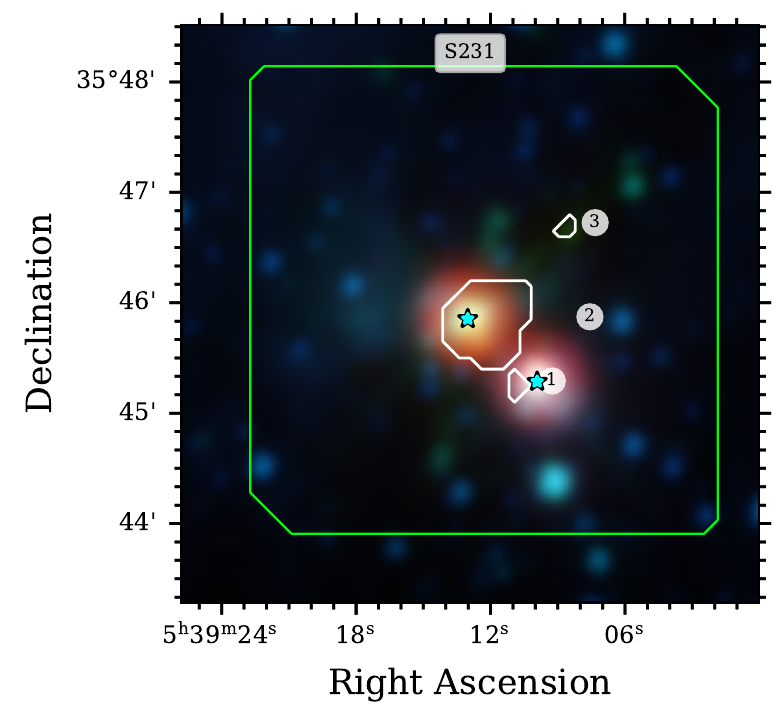}
    \end{minipage}   
    \begin{minipage}{0.45\linewidth}
        \centering \includegraphics[width=\linewidth]{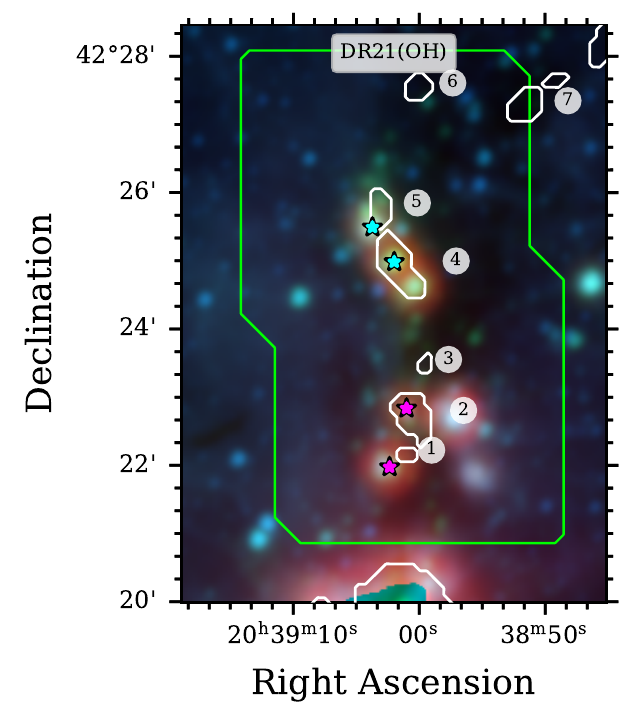}
    \end{minipage}\hfill
    \begin{minipage}{0.45\linewidth}
        \centering \includegraphics[width=\linewidth]{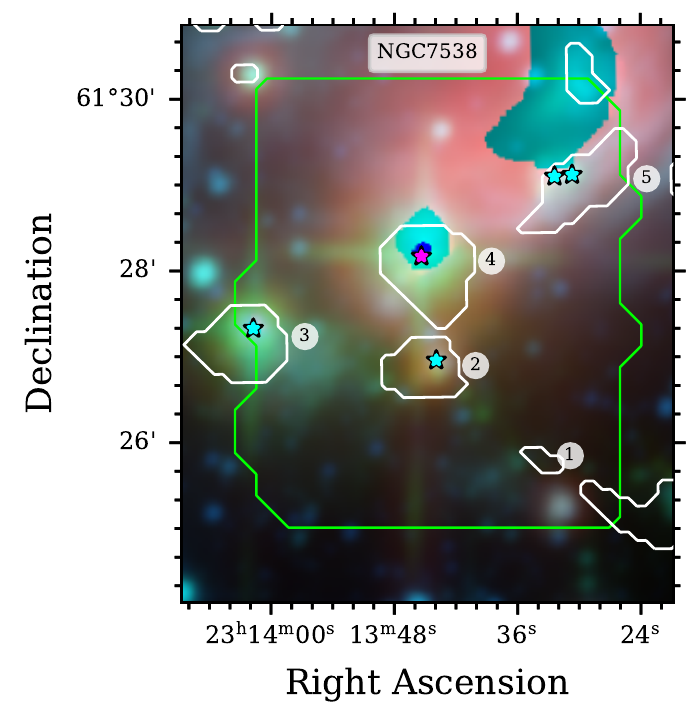}
    \end{minipage}\hfill
   \caption{Wide-field Infrared Survey Explorer (WISE)~\cite{Wright2010} maps at 3.4\micron~(Blue), 4.6\micron~(Green), and 22\micron~(Red). The white contours illustrate the derived clumps. The clump ids are the same as in Fig.\ref{fig:clump}. The cyan and magenta stars indicate YSO and \Hii\ regions, respectively.} \label{fig:wise}
\end{figure}

\FloatBarrier
\section{Table of molecular abundances relative to H$_2$}
\FloatBarrier
      \begin{sidewaystable}
\scriptsize
\setcaptionmargin{0mm}
\onelinecaptionstrue
\captionstyle{flushleft} 
\caption{The molecular abundances relative to H$_2$}
\label{tab:abun}
\bigskip
\begin{tabular}{l|c|c|c|c|c|c|c|c|c|c|c}

\hline
Source & id & HCN & HNC & HCO$^+$ & CS & SO & SO$_2$ &  OCS & HC$_3$N & HNCO & SiO  \\ 
& & $\times10^{-10}$ & $\times10^{-10}$ & $\times10^{-10}$ & $\times10^{-10}$ & $\times10^{-10}$ & $\times10^{-10}$ & $\times10^{-10}$ & $\times10^{-10}$ & $\times10^{-10}$ & $\times10^{-10}$ \\ 
\hline
\multicolumn{12}{c}{YSO} \\ 
\hline 
L1287 & 3 & 702.2 (53.1) & 387.3 (24.1) & 684.3 (16.9) & 138.6 (15.1) & 82.4 (9.6) & 77.1 (18.1) & 51.5 (18.1) & 22.6 (1.4) & 23.5 (2.5) & 9.7 (0.9)
\\
S187 & 1 & 460.2 (73.4) & 198.8 (39.0) & 156.2 (24.4) & 220.8 (26.2) & 59.2 (14.2) & 5.7 (2.4) & 42.2 (21.0) & 32.4 (2.8) & 33.2 (9.3) & --
\\
 & 2 & 270.9 (68.6) & 225.1 (41.4) & 544.4 (59.0) & 85.6 (19.0) & 96.8 (19.9) & 42.0 (22.9) & 58.0 (33.4) & 8.7 (2.5) & 15.6 (3.9) & 6.4 (1.5)
\\
S231 & 1 & 466.1 (39.3) & 92.9 (11.0) & 236.5 (19.1) & 80.5 (7.4) & 64.4 (6.4) & 80.2 (12.6) & 10.4 (4.8) & 12.1 (0.9) & 14.4 (2.3) & 5.2 (0.5)
\\
 & 2 & 431.6 (24.6) & 132.9 (12.0) & 217.2 (10.8) & 91.1 (5.8) & 61.8 (5.4) & 49.8 (10.3) & 29.9 (9.1) & 11.3 (0.6) & 10.0 (1.3) & 7.9 (0.4)
\\
DR~21(OH) & 4 & 327.3 (24.5) & 185.2 (6.9) & 266.3 (9.2) & 109.9 (4.4) & 31.0 (1.6) & 17.4 (3.1) & 18.1 (3.8) & 13.9 (0.5) & 6.4 (0.4) & 1.3 (0.1)
\\
 & 5 & 182.2 (14.8) & 148.3 (8.4) & 220.1 (12.1) & 70.1 (4.2) & 26.4 (1.9) & 32.1 (4.9) & 16.4 (3.1) & 9.2 (0.5) & 5.6 (0.4) & 4.3 (0.3)
\\
NGC~7538 & 2 & 695.2 (20.9) & 151.5 (5.6) & 197.3 (6.0) & 177.3 (5.3) & 88.5 (2.7) & 63.3 (3.0) & 38.7 (2.5) & 13.8 (0.4) & 14.4 (0.5) & 7.7 (0.3)
\\
 & 3 & 929.1 (52.9) & 178.9 (28.8) & 211.4 (18.2) & 147.1 (14.8) & 86.5 (9.0) & 59.2 (17.4) & 19.5 (8.1) & 15.6 (0.7) & 22.7 (3.3) & 4.1 (0.8)
\\
 & 5 & 523.8 (92.7) & -- & 152.0 (24.5) & 117.8 (21.6) & 37.2 (8.0) & 8.0 (3.4) & 54.9 (31.4) & 10.8 (2.9) & 25.7 (5.3) & 1.9 (0.3)\\
\hline 
\multicolumn{12}{c}{\Hii\ region} \\ 
\hline 
DR~21(OH) & 1 & 279.3 (19.2) & 113.4 (7.0) & 124.4 (7.7) & 97.3 (6.1) & 35.1 (2.2) & -- & 12.6 (1.4) & 10.6 (0.7) & 4.8 (0.3) & 2.2 (0.1)
\\
 & 2 & 123.5 (20.8) & 60.5 (4.7) & 77.3 (13.0) & 39.2 (9.6) & 13.3 (3.0) & 49.9 (12.6) & 9.8 (1.8) & 6.4 (0.6) & 3.8 (0.3) & 2.0 (0.3)
\\
NGC~7538 & 4 & 653.6 (16.7) & 78.4 (6.2) & 143.0 (4.7) & 173.4 (4.6) & 55.9 (2.3) & 40.1 (4.9) & 18.8 (3.3) & 6.0 (0.2) & 8.2 (1.0) & 4.2 (0.2)\\ 
\hline 
\multicolumn{12}{c}{submm} \\ 
\hline 
L1287 & 1 & 457.3 (34.3) & 374.5 (20.3) & 531.5 (18.5) & 115.4 (12.6) & 69.8 (8.8) & 55.8 (16.7) & 83.0 (19.1) & 11.6 (0.8) & 16.2 (1.6) & 2.5 (0.5)
\\
 & 2 & 364.6 (61.8) & 395.0 (38.1) & 471.9 (31.5) & -- & -- & 15.5 (nan) & 73.7 (47.3) & 12.9 (1.4) & 11.5 (1.8) & --
\\
S231 & 3 & 341.2 (39.8) & 104.5 (16.5) & 354.4 (29.0) & 81.5 (9.4) & 162.7 (13.5) & 129.3 (13.4) & 55.8 (10.1) & 3.0 (0.4) & 20.7 (2.3) & 4.4 (0.4)
\\
DR~21(OH) & 3 & 202.0 (15.1) & 136.5 (9.3) & 155.7 (12.0) & 91.5 (6.0) & 39.6 (2.6) & 80.1 (14.9) & -- & 7.1 (0.7) & 8.6 (0.6) & 1.6 (0.2)
\\
 & 6 & 169.8 (22.7) & 423.3 (44.4) & 645.7 (41.1) & 124.1 (22.7) & 92.8 (18.4) & 144.9 (35.1) & -- & 7.9 (1.2) & 22.0 (4.4) & --
\\
 & 7 & -- & 711.9 (129.6) & 965.6 (100.4) & -- & 210.2 (37.7) & -- & -- & 21.2 (4.9) & -- & --
\\
NGC~7538 & 1 & 736.6 (51.2) & 264.3 (25.7) & 437.2 (26.4) & 160.3 (16.2) & 31.7 (5.2) & 85.9 (19.2) & 15.0 (8.6) & 10.9 (0.9) & 8.5 (1.6) & 2.7 (0.5) \\
\hline
\multicolumn{12}{l}{The errors for each parameter are shown in parentheses.}

    \end{tabular}
 \end{sidewaystable}

\end{appendix}


\end{document}